\documentclass[a4paper,11pt]{article}
\usepackage{jinstpub} 

\usepackage{orcidlink} 

\usepackage{hyperref}
\hypersetup{colorlinks,breaklinks,
           citecolor=[RGB]{0,96,0},
           linkcolor=[RGB]{0,0,0},
           urlcolor=[RGB]{0,0,96}}

\usepackage{multirow}
\usepackage{subcaption}
\usepackage{threeparttable}
\usepackage{pifont} 

\usepackage{siunitx}
\DeclareSIUnit{\electronvolt}{eV}
\DeclareSIUnit{\pixel}{px}
\newcommand{\var}[2]{#1_{\mathrm{#2}}} 

\title{A high-flux electron detection system to measure non-linear Compton scattering at LUXE}

\author[a,b]{Antonios~Athanassiadis\orcidlink{0009-0008-9963-0024}}
\author[c]{Oleksandr~Borysov\orcidlink{0000-0002-5384-611X}}
\author[a]{Florian~Burkart\orcidlink{0000-0002-2178-4639}}
\author[a]{Hannes~Dinter-Habermeier}
\author[a]{John~Hallford\orcidlink{0000-0001-7159-4078}}
\author[a,b]{Beate~Heinemann\orcidlink{0000-0002-1673-7926}}
\author[a]{Louis~Helary\orcidlink{0000-0001-7891-8354}}
\author[a,d]{Luke~Hendriks\orcidlink{0009-0000-0263-5980}}
\author[a]{Ruth~Jacobs\orcidlink{0000-0001-5446-5901}}
\author[a]{Max~Kellermeier\orcidlink{0000-0002-8840-782X}}
\author[a]{Willi~Kuropka\orcidlink{0000-0002-8649-7366}}
\author[a]{Jenny~List\orcidlink{0000-0002-0626-3093}}
\author[a]{Frank~Mayet\orcidlink{0000-0003-3816-0686}}
\author[b]{Gudrid~Moortgat-Pick\orcidlink{0000-0002-5145-901X}}
\author[a]{Evan~Ranken\orcidlink{0000-0001-7472-5029}}
\author[a]{Stefan~Schmitt}
\author[a,e,1]{Ivo~Schulthess\orcidlink{0000-0002-5621-2462}\note{Corresponding author}}
\author[a]{Thomas~Vinatier}
\author[a,d]{Matthew~Wing\orcidlink{0000-0002-6319-4135}}

\affiliation[a]{Deutsches Elektronen-Synchrotron DESY, 22603 Hamburg, Germany}
\affiliation[b]{Universität Hamburg, 20148 Hamburg, Germany}
\affiliation[c]{Weizmann Institute of Science, 7610001 Rehovot, Israel}
\affiliation[d]{University College London, London WC1E 6BT, United Kingdom}
\affiliation[e]{Institute for Particle Physics and Astrophysics, ETH Zurich, 8093 Zurich, Switzerland}

\emailAdd{ivo.schulthess@desy.de}

\abstract{This paper presents the development and testing of a high-flux electron detection system designed to measure the energy spectrum of more than $10^9$ electrons from laser--electron collisions at the LUXE experiment, planned at DESY. The system is intended to enable precision measurements of non-linear Compton scattering in the strong-field regime of quantum electrodynamics. It combines a scintillating screen observed by a camera system with a spatially segmented detector based on the Cherenkov effect. Prototype tests were conducted at two accelerator facilities, where the performance of the system was evaluated under realistic beam conditions. The results demonstrate the feasibility of the detector concept, characterize its current performance and limitations, and provide input for the optimization and final design of the detection system to be used in the experiment. }

\notoc

\begin{document}

\maketitle
\flushbottom

\section{Introduction}\label{sec:intro}

Strong-field quantum electrodynamics (QED) describes the interactions of charged particles and photons in the presence of intense electromagnetic fields, where non-linear quantum effects become relevant~\cite{Ritus:1985vta, DiPiazza:2011tq, Gonoskov:2021hwf, Fedotov:2022ely, Kropf:2025loq}. These phenomena become pronounced when the field strength approaches the QED critical field, or Schwinger limit, defined as $E_\mathrm{qed} = m_e^2 c^3/(e \hbar) \approx \SI{1.3e18}{V/m}$, with $m_e$ denoting the electron mass, $c$ the speed of light in vacuum, $e$ the elementary charge, and $\hbar$ the reduced Planck constant. In this regime, phenomena such as non-linear Compton scattering and non-linear Breit--Wheeler pair production occur. Recent advances in high-intensity laser technology have allowed exploration of this regime in a laboratory environment~\cite{Danson:2019qlu}.

Strong-field QED phenomena can occur naturally in various environments, most commonly associated with extreme astrophysical objects like magnetars, neutron stars, and black holes, where magnetic fields in excess of the Schwinger limit may lead to strong-field QED effects~\cite{Kim:2025uhi}. They can also appear in the strong electric fields arising from the Coulomb potential in crystals~\cite{lindhard_motion_1964, Uggerhj:2005wgn, DiPiazza:2019vwb}, a scenario being investigated by the NA63 experiment at CERN~\cite{Nielsen:2023icv}. In high-energy electron--positron colliders, beam particles radiate photons because of the interaction with the electromagnetic fields produced by the opposite beam~\cite{Yokoya:1991qz, Yokoya:2000bv}. This effect is called beamstrahlung and becomes of particular importance for future lepton colliders~\cite{Schulte:1999xb, Barklow:2023iav, LinearColliderVision:2025hlt}. At the highest envisaged beam energies, the quantum beamstrahlung regime can be reached, which cannot be reliably modeled with today's simulation tools, making the study of strong-field QED particularly necessary and timely. 

Fundamental tests of the theory can be performed in electron--laser or photon--laser collisions. At SLAC, strong-field QED has been studied via electron--laser collisions at the E-144 experiment~\cite{Burke:1997ew}, and such collisions are now being pushed even further into the non-perturbative regime with the E-320 experiment at the FACET-II facility~\cite{reisE320ProgressFY242024, Athanassiadis:2025git}. Significant efforts are being made to test strong-field QED in plasma experiments, albeit often with substantial uncertainties~\cite{Cole:2017zca}. At DESY Hamburg, the Laser Und XFEL Experiment (LUXE)~\cite{Abramowicz:2021zja, LUXE:2023crk} plans to use the electron beam of the European XFEL~\cite{Altarelli:2006zza}. For the parameter space tested at LUXE, the Monte Carlo tool Ptarmigan~\cite{Blackburn:2023mlo, ptarmigan_github_2024} was developed to simulate the interaction between high-energy particle beams and intense laser pulses. The tool includes the classical dynamics and strong-field QED processes. 

\begin{figure}[tbh]
    \centering
    \includegraphics[width=0.9\linewidth]{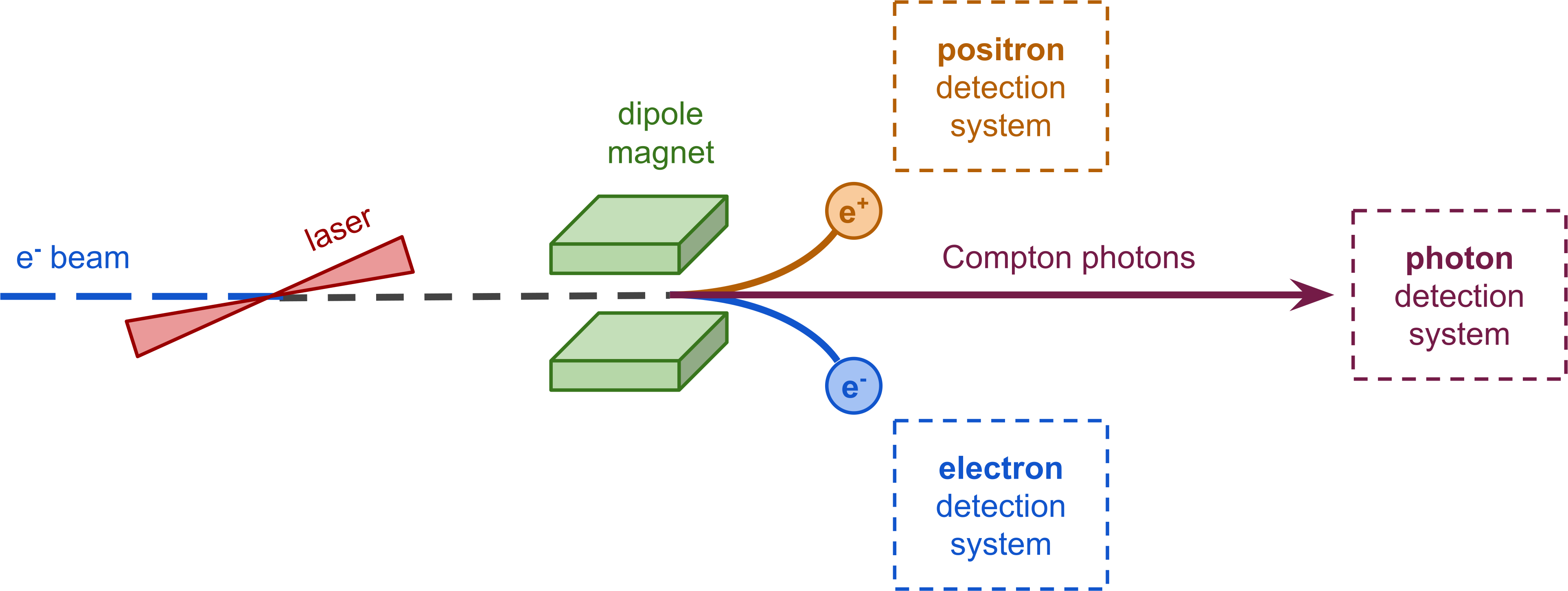}
    \caption{Schematic of the LUXE experiment in the electron--laser mode. The electron beam of the European XFEL (blue) will collide with a high-intensity laser pulse (red). The electrons can scatter via the non-linear Compton process, losing energy through the emission of Compton photons, while electron--positron pairs can be produced via the non-linear Breit--Wheeler process. The electrons (blue) and positrons (orange) will be spatially separated with a dipole magnet (green) and detected with their corresponding detection system. The Compton photons (purple), which are unaffected by the magnet, will also be detected. }
    \label{fig:luxeSchematic}
\end{figure}

A high-flux electron detection system (EDS) is being developed to measure the energy spectrum of electrons produced in laser--electron collisions at the LUXE experiment planned at DESY. The experiment aims to probe the transition from linear QED to non-linear and non-perturbative QED by colliding electrons from the European XFEL with a high-intensity laser.\footnote{An operation mode in which high-energy photons collide with the laser is also planned, where the electrons are first converted into photons that then collide with the laser. However, in this mode the EDS will only be used to monitor the photon production.} A schematic of the experiment is shown in figure~\ref{fig:luxeSchematic}. The accelerator has a repetition rate of \SI{10}{Hz} and provides electrons with an energy of up to \SI{16.5}{\giga\electronvolt} and a bunch charge of $\SI{250}{pC} \approx 1.5 \times 10^9~\var{N}{e}$. A titanium-doped sapphire (Ti:sapphire) laser with a wavelength of \SI{800}{\nano\meter} and a peak power of \SI{40}{\tera\watt} is planned in the initial phase-0, with an upgrade to \SI{350}{\tera\watt} foreseen in phase-1. This will allow access to peak intensities of up to \SI{1.3e20}{\watt\per\centi\meter\squared} (phase-0) and \SI{12e20}{\watt\per\centi\meter\squared} (phase-1), corresponding to laser intensity parameters $a_0$ of \num{7.9} and \num{23.6}, respectively~\cite{Abramowicz:2021zja, LUXE:2023crk}. The laser has a repetition rate of \SIrange{1}{10}{\hertz} which defines the collision rate. In case the laser is operated at a lower repetition rate than the electron beam, the other electron bunches can be used for background characterization. Some details of the LUXE setup are dependent on the precise installation location, which is currently being finalized. This paper uses the proposed setup as laid out in the technical design report~\cite{LUXE:2023crk}. 

The main purpose of the EDS is to characterize the Compton spectrum of the electrons resulting from the strong-field QED interaction. LUXE foresees a combination of two complementary methods for each detection system in order to minimize uncertainties and allow for cross-calibrations.  The envisioned design of the LUXE EDS is described in section~\ref{sec:eds}, together with the concept of the detector. In section~\ref{sec:detectorTests} we describe the tests that were performed with two prototypes at the ARES and FACET-II facilities. Finally, the results are summarized in section~\ref{sec:conclusion}.

\section{The Electron Detection System}\label{sec:eds}

\begin{figure}[tbh]
    \centering
    \includegraphics[width=0.9\linewidth]{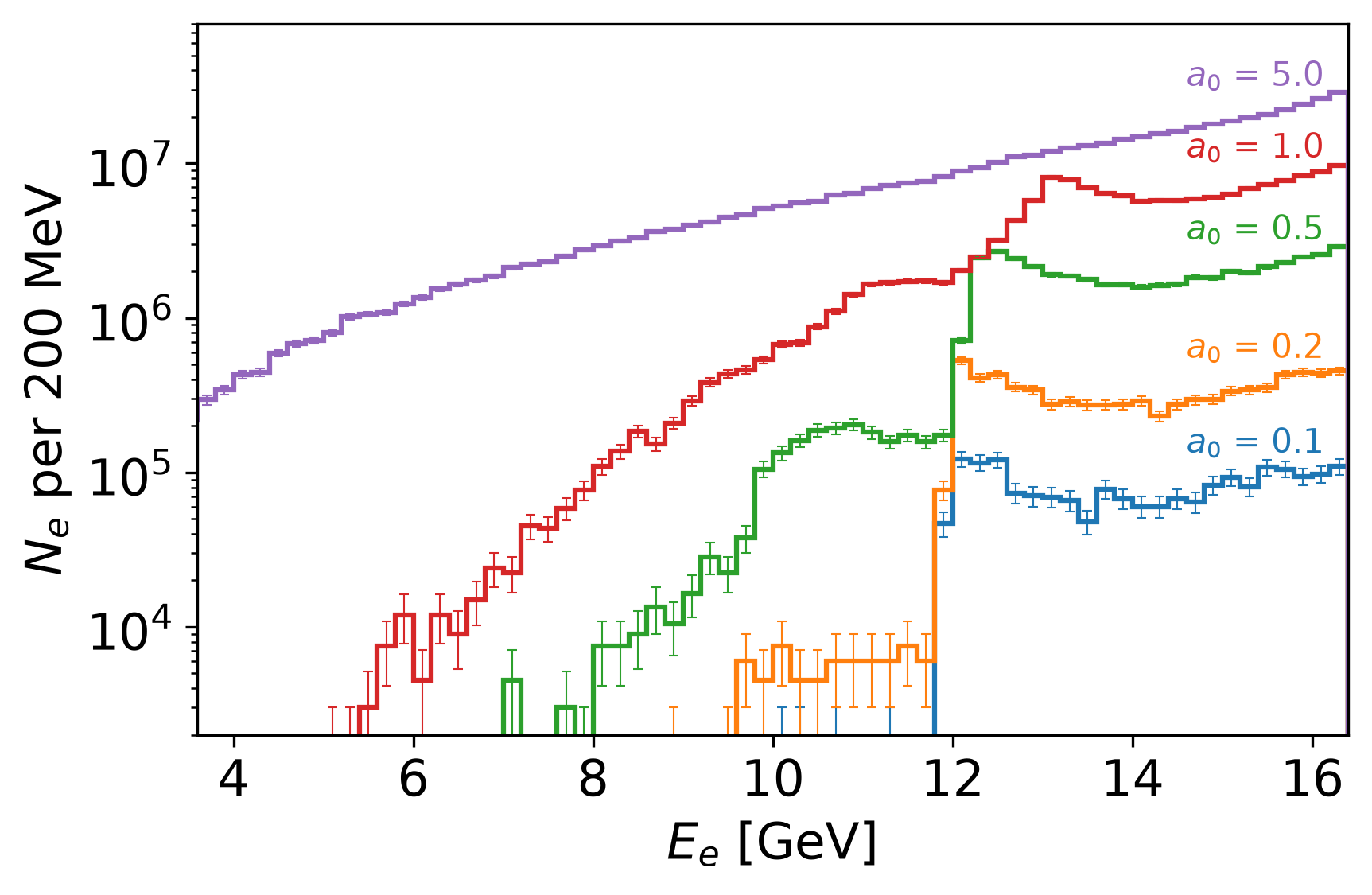}
    \caption{Simulations of electron--laser collisions for different laser intensity parameters $a_0$. The number of scattered Compton electrons per bunch crossing is given with a bin size of \SI{200}{\mega\electronvolt}. The error bars indicate the statistical uncertainties of the simulated event samples. The blue line ($a_0 = 0.1$) corresponds to the case of linear Compton scattering. The non-interacting electrons from the main electron beam with an energy of \SI{16.5}{GeV} are excluded~\cite{Athanassiadis:2025git}. }
    \label{fig:electronComptonSpectra}
\end{figure}

An accurate measurement of electron energy distributions as a function of laser intensity is crucial for validating theoretical models in the strong-field QED domain at the parameters expected at LUXE. To measure the electron energy spectrum, the LUXE setup includes a \SI{1.2}{\meter}-long dipole magnet with a maximum field strength of \SI{1.6}{\tesla} that separates charged particles spatially with respect to their energy. The EDS is placed \SI{2.8}{\meter} behind the dipole magnet. Electron fluxes on the order of $10^3$ to $10^7$ electrons per bunch crossing are expected within an energy interval of \SI{200}{\mega\electronvolt}, requiring a detector system with a large dynamic range in sensitivity. A Ptarmigan simulation presented in figure~\ref{fig:electronComptonSpectra} shows the electron energy distribution after the electron--laser interaction. At the location of the linear Compton edge at \SI{12}{\giga\electronvolt}, an energy interval of \SI{200}{\mega\electronvolt} corresponds to a spatial separation of approximately \SI{3}{\milli\meter} at the EDS when operating the dipole magnet at its maximum field strength. 

Besides the measurement requirements, the EDS faces numerous other challenges. A detailed Geant4-based simulation of the extraction beamline from the European XFEL and the experimental setup of LUXE including beam dumps has estimated the ionizing dose experienced by the EDS to be up to \SI{2e4}{Gy} over an annual operation of about \SI{1e7}{\second}~\cite{Abramowicz:2021zja}. Moreover, electromagnetic pulses generated by accelerator components and high‑power laser systems introduce additional sources of background noise and potential interference. 

\begin{figure}[tbh]
    \centering
    \includegraphics[width=0.7\linewidth]{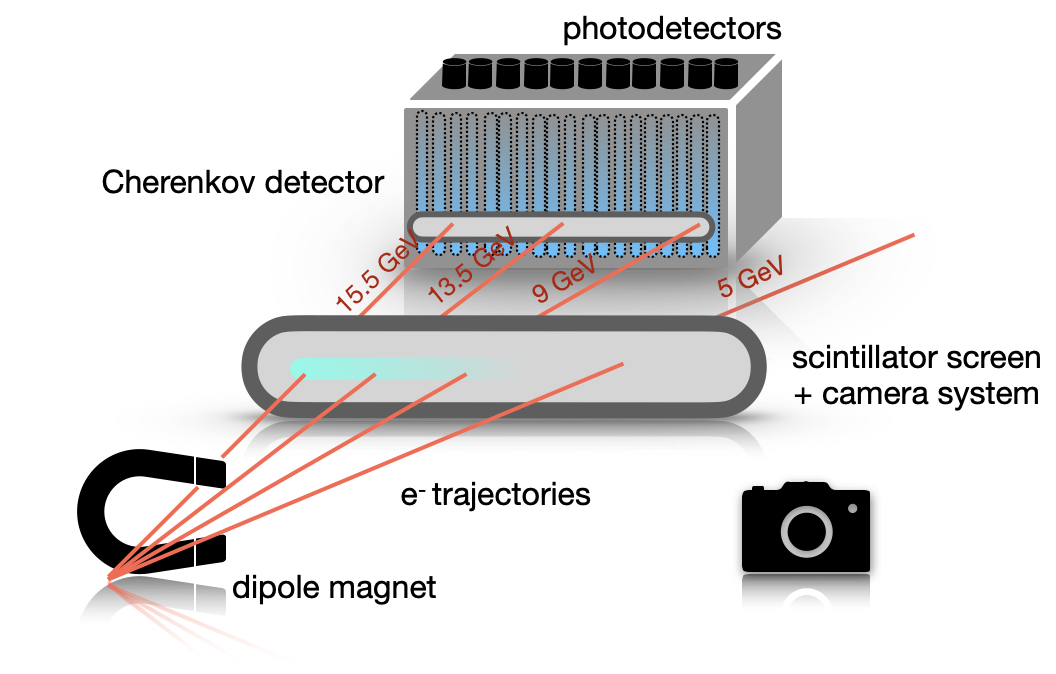}
    \caption{Schematic of the EDS at LUXE. Located downstream of the spectrometer dipole, the primary electron beam and the produced secondary particles are spatially separated by energy. The electron energy spectrum can be obtained by evaluating the flux of these spatially distributed electrons using the scintillating screen with a camera system, as well as the Cherenkov-based straw detector~\cite{LUXE:2023crk}.}
    \label{fig:sketch_luxe_eds}
\end{figure}

To fulfill the measurement requirements while operating under these challenging conditions, the EDS comprises two complementary detector technologies for precise cross-calibration, spatial alignment, and the assessment of systematic effects: a scintillating screen observed with an optical camera system, followed by a spatially segmented Cherenkov detector which sits \SI{0.4}{\meter} behind the screen (figure~\ref{fig:sketch_luxe_eds}). The screen--camera system delivers a simple solution to measure electron spectra with a final position resolution of \SI{500}{\micro\meter}, while the Cherenkov detector has been designed to ensure a relative energy resolution of less than 2\% using a channel spacing of \SI{3}{\milli\meter}~\cite{Abramowicz:2021zja}. The following sections discuss the characteristics of the proposed detector system, including its various components and their operating principles.

\subsection{Screen Detector}\label{sec:screen}

The screen detector of the EDS at LUXE will consist of a scintillating screen, with three CMOS~\cite{stefanov_cmos_2022} cameras and their corresponding optics. The scintillating material, terbium-doped gadolinium oxysulfide, Gd$_2$O$_2$S:Tb (GadOx), was chosen for its high light yield, radiation resistance, and wavelength profile \cite{LUXE:2023crk}. The scintillation light created by a GadOx screen has a main emission peak around \SI{543}{nm}. Similar to other inorganic scintillators, this emission is produced by the de-excitation of the activator ion (here: Tb$^{3+}$). The Tb$^{3+}$ ions are excited to a higher energy state due to the transfer of energy from the Gd$^{3+}$ ions by electron--hole pairs created by incident particles \cite{HERNANDEZADAME20188}.

The energy deposited in the scintillating material determines the number of electron--hole pairs created and thus the light yield. For relativistic electrons, the light yield of GadOx-based scintillating screens depends only weakly on the incident electron energy~\cite{glinec_absolute_2006, Nakamura:2011zzc}. Consequently, the total scintillation light yield can be used to determine the number of electrons incident on the screen, provided that the energy dependence of the response is accounted for in the detector calibration. Nonetheless, the screen will also be sensitive to low-energy background particles, as in this regime, the factor 1/$\beta^2$ dominates. It is therefore essential to minimize any background radiation. Furthermore, impurities on the screen should be minimal so that the screen response is independent of the location of an electron hit. 

For the EDS at LUXE, the dimensions of the scintillating screen are chosen to be \SI{500}{mm} along the energy axis by \SI{100}{mm} in the vertical direction. Taking into account the preceding dipole magnet, the screen will be able to measure electrons with an energy between \SI{3.1}{GeV} and \SI{15}{GeV}. The total thickness of the screen will be around \SI{0.5}{mm}; in radiation lengths, this is $\sim 2.8\%$, which makes the screen thin enough to minimally disturb transiting electrons before they reach the Cherenkov detector. 

The three cameras will record different parts of the scintillating screen. Two cameras of type \mbox{Basler acA1920-40gm}~\cite{2KCam} with a resolution of $1920\times\SI{1200}{pixels}$ (2K-camera), will each observe half of the screen, while a third camera (\mbox{Basler acA4096-11gm \cite{4KCam}}) with a resolution of $4096\times\SI{2160}{pixels}$ (4K-camera) will cover the crucial high-energy region of about \SI{212}{\milli\meter}, where the first Compton edge is visible. These particular cameras were selected because they provide a high photon detection efficiency (approximately $70\%$) near the main emission peak of the GadOx screen, as well as a linear response in light yield. Due to the magnetic spectroscopy mechanism, the electrons at higher energies will be spatially less separated, and thus, a camera with a higher resolution is needed in this region of the Compton spectrum. Additionally, the finer resolution combined with a lens with a higher focal length leads to a lower signal per pixel. This effectively increases the dynamic range of the combined camera system, since the 4K-camera saturates at a higher photon flux. The optics of the cameras include fixed-focal-length lenses and optical filters. The 2K-cameras will both have lenses with a focal length of \SI{50}{mm}, while the 4K-camera will have a lens  with a focal length of \SI{75}{mm}. For the proposed LUXE setup, this leads to a nominal positional resolution of \SI{135.9}{\upmu m} and \SI{51.8}{\upmu m} for the 2K and 4K-cameras, respectively. Optical filters that accept a narrow band of wavelengths will be used, centered around \SI{543}{nm}, the emission peak of the GadOx screen. This reduces the effect of ambient light on the measurements. Further details of the proposed scintillating screen detector setup at LUXE are provided in~\cite{LUXE:2023crk}.

\subsection{Straw Detector}\label{sec:straws}

The second detector technology of the EDS is based on the Cherenkov effect. When charged particles pass through a medium with a phase velocity larger than the speed of light in this medium, Cherenkov photons are emitted. For relativistic electrons well above the Cherenkov threshold, the number of photons $\var{N}{Ch}$ emitted per electron is almost independent of the electron's energy and can be calculated using the Frank-Tamm formula~\cite{Leroy:2011goz}
\begin{align}\label{eq:n_cherenkov}
    \var{N}{Ch}(L,\var{n}{m}) &\approx \kappa \times L \times \biggl(1-\frac{1}{(\beta\var{n}{m})^2}\biggr), \quad \text{with}~\kappa = 2\pi\alpha z^2 \biggl(\frac{1}{\lambda_1}-\frac{1}{\lambda_2}\biggr) = \mathrm{const.} \, ,
\end{align}
where $L$ is the particle track length in medium, $\var{\lambda}{1,2}$ the considered min./max.\ photon spectrum wavelength, $\alpha$ the fine structure constant, $z$ the particle charge number, and $\var{n}{m}(\lambda) \approx \mathrm{const.}$ the refractive index of the material. 

In the EDS prototypes, air-filled stainless-steel straws ($\var{n}{air} \approx 1.00028$) and solid glass rods ($\var{n}{glass} \approx 1.53$) were used. The refractive index determines both the Cherenkov threshold and the photon yield. For the two media, the corresponding electron kinetic-energy thresholds are approximately \SI{21}{\mega\electronvolt} in air and \SI{0.16}{\mega\electronvolt} in glass. Below these thresholds, no Cherenkov light is emitted, providing an intrinsic suppression of low-energy charged-particle backgrounds. Above the threshold, a charged particle traveling through a medium with a low refractive index will create fewer photons than in a medium with a higher refractive index. Using equation~\eqref{eq:n_cherenkov}, the Cherenkov photon yield per unit track length in air is approximately three orders of magnitude lower than in glass. Hence, the sensitivity of channels using the glass rods is expected to be significantly increased. The performance of the Cherenkov detector prototypes was characterized using primary electron beams at accelerator facilities, as described in sections~\ref{sec:ares} and~\ref{sec:facet}. For these measurements, the electron beam can be approximated by a Gaussian transverse profile. Since many electrons pass through a straw, the number of created Cherenkov photons can be calculated from the combined track length of all electrons, $\var{L}{total}$. This track length varies with respect to the position due to the round shape of the straw. To account for this effect, the number of created photons by a Gaussian-shaped electron bunch $G(\mu,\sigma;x)$ can be derived numerically as 
\begin{align}
    \var{N}{total} &= \var{N}{Ch} (\var{L}{total}(Q, r, \sigma; \mu), \var{n}{m}) \, ,\label{eq:n_total} \\ 
    \nonumber \\
    \var{L}{total}(Q,r,\sigma;\mu) \, \biggl|_{Q,r,\sigma} &= 2 Q \int_{-r}^{+r} \sqrt{r^2-x^2} \times G(\mu,\sigma;x) \, \mathrm{d}x \, ,
    \label{eq:l_total}
\end{align}
where $\mu$ is the distance of the Gaussian beam center to the straw center at $x = 0$, $\sigma$ is its width, $Q$ is the bunch charge, and $r$ is the straw radius. 

\begin{figure}[tbh]
    \centering
    \begin{subfigure}[t]{0.49\textwidth}
        \centering
        \includegraphics[width=0.7\linewidth]{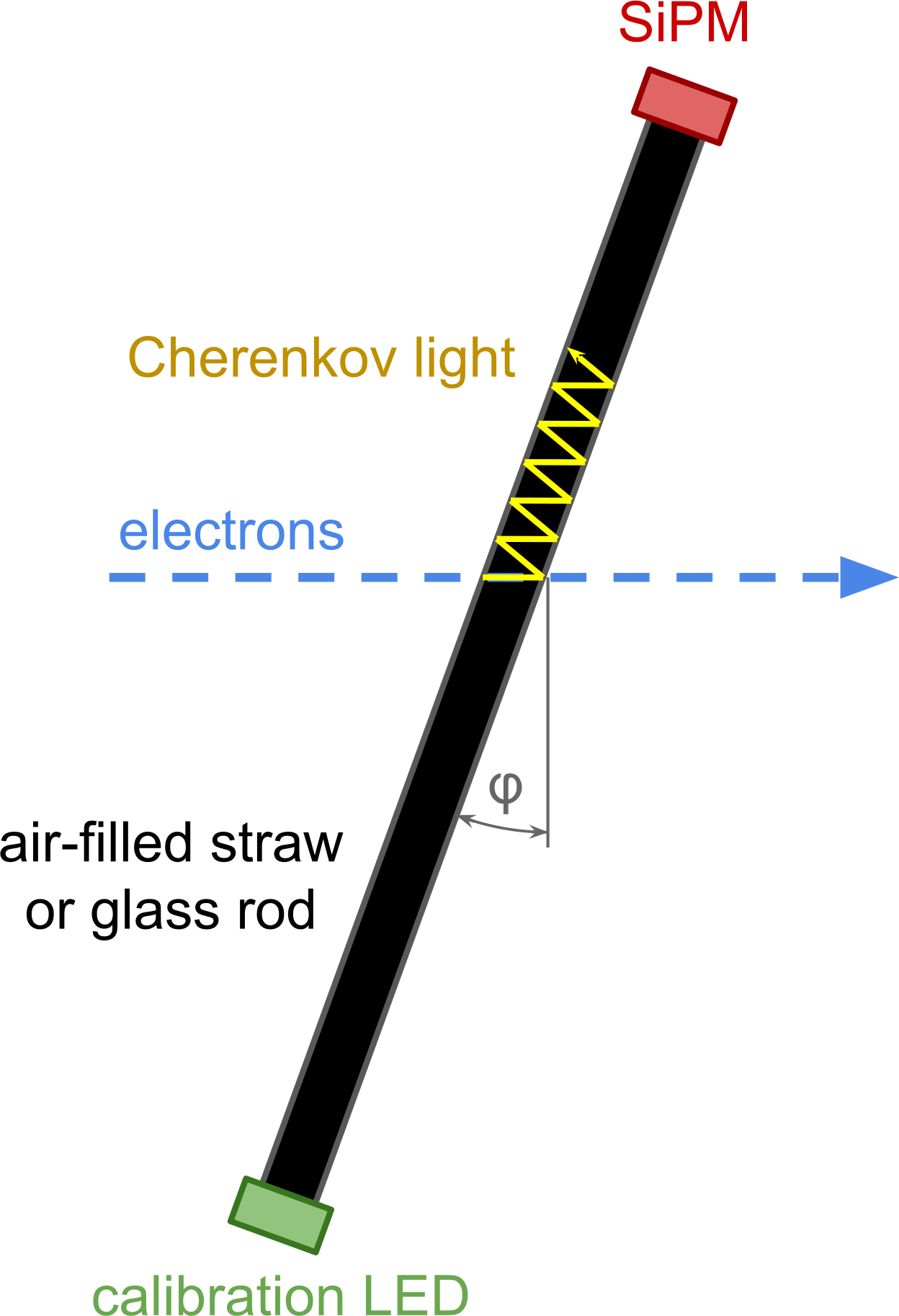}
        \caption{}
        \label{fig:straw_concept}
    \end{subfigure}
    \hfill 
    \begin{subfigure}[t]{0.49\textwidth}
        \centering
        \includegraphics[width=\linewidth]{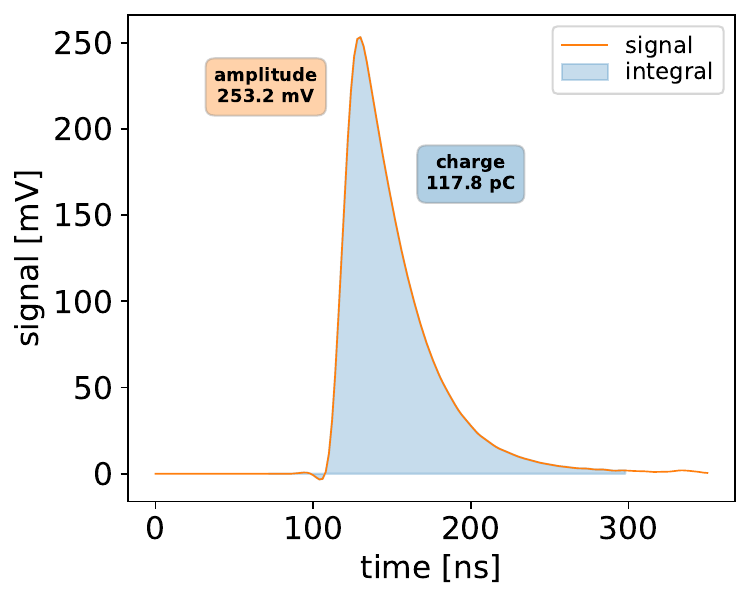}
        \caption{}
        \label{fig:led_pulse}
    \end{subfigure}
    \caption{a) Schematics of a relativistic electron (blue) passing through an air-filled straw or a glass rod (black). The Cherenkov light (yellow) is emitted at a medium-dependent angle with respect to the electron trajectory, which is not represented in the schematic, and is reflected at the surrounding surfaces towards a SiPM (red) sitting at the upper end. On the opposite side, an LED- and optical fiber-based calibration system (green) is placed. The straw or rod can be tilted by angle $\varphi$~\cite{Abramowicz:2021zja}. b) Electrical signal response of a SiPM recording an LED light pulse. The signal intensity is measured as the total charge collected.}
    \label{fig:eds_detectors}
\end{figure}

To efficiently transport the generated Cherenkov photons to the photosensor, the inner surface of the stainless-steel straws is polished, while the glass rods are wrapped in a reflective foil. When an electron passes through such a straw or rod, the Cherenkov light created along the transit through the material is emitted at an angle with respect to the electron trajectory, reflected by the inner surfaces and guided towards the photosensor at one of the ends, as illustrated in figure~\ref{fig:straw_concept}. The tilt angle $\varphi$ with respect to the orientation perpendicular to the beam axis can be used to tune the number of photons reaching the photosensor by altering the number of reflections. The opposite end of the straw holds an LED that can be pulsed and therefore serves as a calibration and monitoring system for the silicon photomultiplier (SiPM), the type of photosensors used in the prototype detectors. 

The SiPMs were electronically connected to a CAEN digitizer~\cite{caenspa_ds3153_2019, caenspa_ds3159_2019} controlled using a modified DAQ software for CAEN digitizers based on~\cite{dejong_caenv1730daq_2023}. The collected SiPM charge, proportional to the waveform integral, is used for the analysis since it is less prone to electronic noise than the peak amplitude.

Lastly, the calibration system allowed UV-LED pulses with a pulse length of about \SI{1}{\nano\second} and a wavelength of \SI{385}{\nano\meter} to be injected from a single LED into the straws via a fiber bundle. An example of such a signal is shown in figure~\ref{fig:led_pulse}. The LED signals can be used to normalize and calibrate individual SiPM signal channels and to determine signal changes over time. Since the light yield that reaches each individual SiPM channel varies due to mechanical and electrical differences, it is only possible to measure changes relative to an initial reference measurement.

\section{Detector Tests}\label{sec:detectorTests}
 
\begin{table}[b]
    \centering
    \begin{tabular}{l || c | c | c | c}
        ~ & ARES~\cite{Burkart:2022kdx} & FACET-II~\cite{Yakimenko:2019sya} & E-320~\cite{reisE320ProgressFY242024} & LUXE phase-0~\cite{Altarelli:2006zza, Abramowicz:2021zja} \\ \hline \hline
        beam energy [GeV] & 0.155 & 10 & 10 & 16.5  \\ \hline
        bunch charge [pC] & \numrange{0.9}{110} & 1400 & 1400 & 250 \\  \hline
        repetition rate [Hz] & 10 & 10 & 10 & 10 \\ \hline \hline
        laser pulse energy [J] & -- & -- & 0.6 & 1.2 \\ \hline
        laser pulse length [fs] & -- & -- & 42 & 30   
    \end{tabular}
    \caption{Key parameters of the facilities or experiments used during EDS test campaigns (ARES, FACET-II, and E-320) and at LUXE, where the EDS is envisioned to operate using electrons from the European XFEL and a \SI{40}{\tera\watt} laser system during phase-0.}
    \label{tab:facilityParameters}
\end{table}

\begin{figure}[tbh]
    \centering
    \begin{subfigure}[t]{0.398\textwidth}
        \centering
        \includegraphics[width=\linewidth]{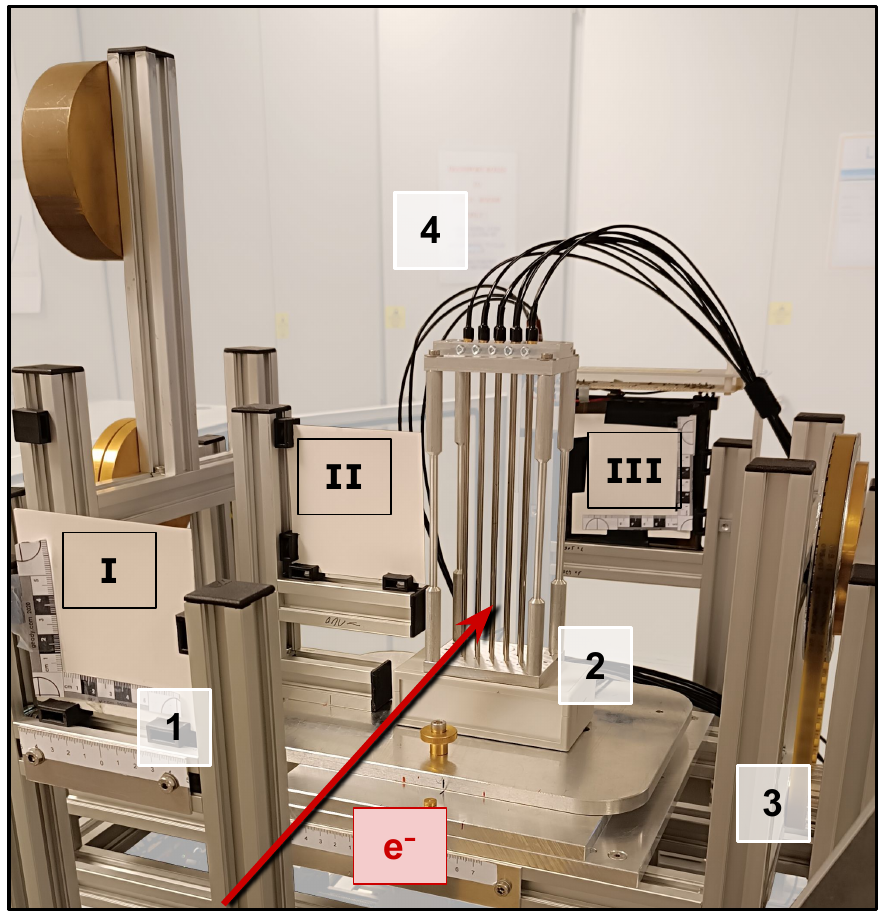}
        \caption{~}
        \label{fig:edsPhotoAres}
    \end{subfigure}
    \hfill
    \begin{subfigure}[t]{0.585\textwidth}
        \centering
        \includegraphics[width=\linewidth]{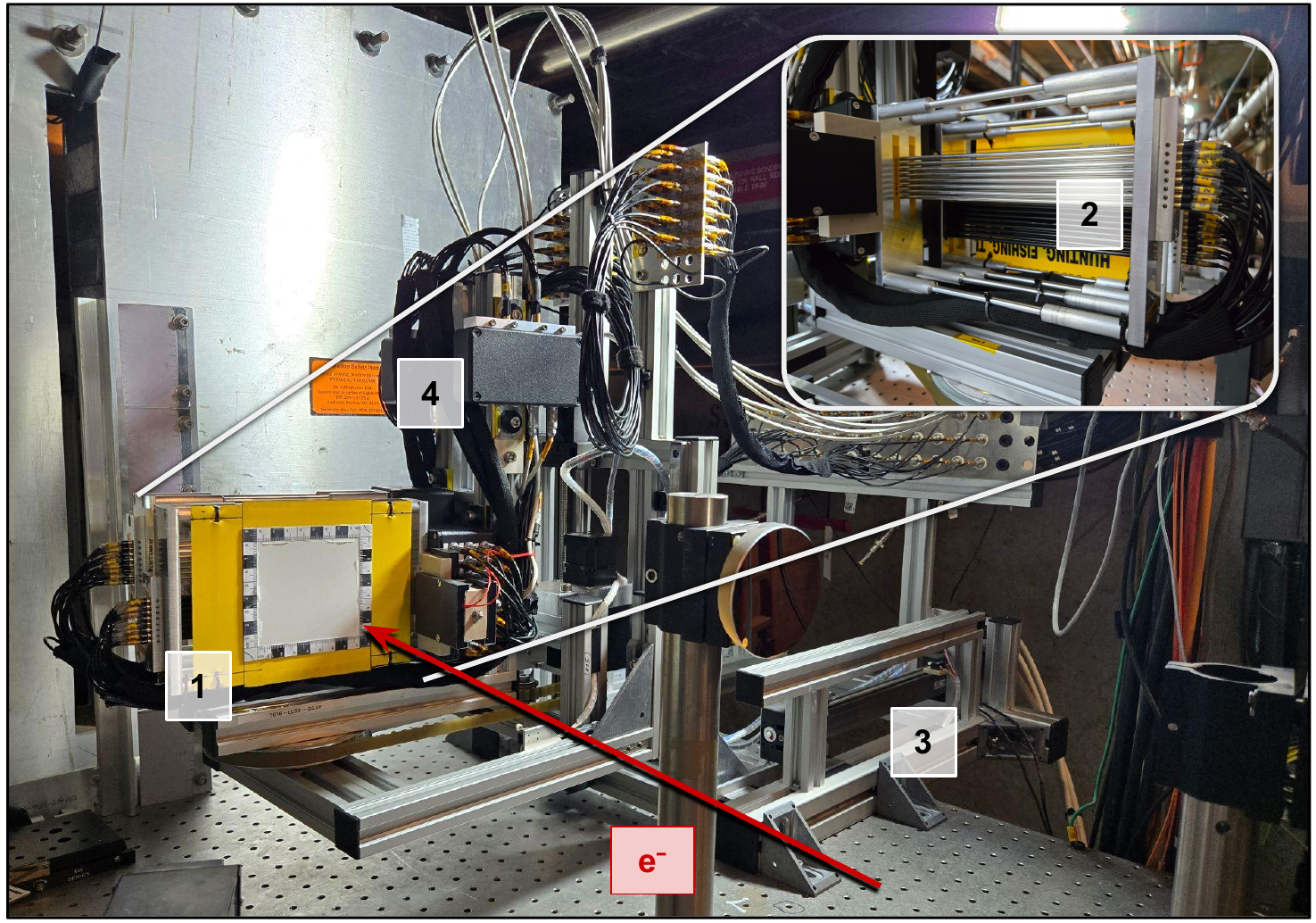}
        \caption{~}
        \label{fig:edsPhotoFacet}
    \end{subfigure}
    \caption{Photograph of the EDS detector prototypes for tests at ARES (a) and at FACET-II (b). The beam direction is indicated by the red arrows. The setup made use of (1) scintillating screens, (2) air-filled steel straws and glass rods, (3) a mechanical frame with rotational and linear stages, and (4) an LED calibration system. At ARES, the three scintillating screen locations are labeled I, II, and III. }
    \label{fig:edsPhotos}
\end{figure}

In order to validate the detector concept and determine the detector's performance, a series of prototype setups have been developed and subjected to testing with electron beams at the Accelerator Research Experiment at SINBAD (ARES) at DESY~\cite{Burkart:2022kdx} and the Facility for Advanced Accelerator Experimental Tests II (FACET-II) at SLAC~\cite{Yakimenko:2019sya}. Two photographs of the latest prototypes used for these studies are shown in figure~\ref{fig:edsPhotos}. 

The measurement campaign at ARES (section~\ref{sec:ares}) served primarily to assess the individual components of the EDS. Different types of scintillating screens and SiPMs were tested and compared under various beam conditions. The measurement campaign at FACET-II (section~\ref{sec:facet}) provided more realistic conditions due to the high energy and bunch charge of the accelerator, resulting also in a harsher radiation environment. It enabled the combined testing of the two detector types comprising the EDS using electrons from laser--electron interactions dispersed by a magnetic spectrometer, resembling the basic measurement concept foreseen at LUXE.

\subsection{ARES}\label{sec:ares}

ARES is a linear accelerator facility with the goal of testing accelerator hardware components, such as injection kickers or bunch compressors, to develop beam diagnostics, including beam position and beam loss monitors, and to explore machine-learning-based accelerator controls. Some of the relevant beam parameters of ARES are given in table~\ref{tab:facilityParameters}~\cite{Burkart:2022kdx}. Because of the capability to change the electron bunch charge, ARES was ideally suited to study the performance of various types of scintillating screens and the straw-based Cherenkov detector concept with the prototype detector system. The detector prototype assembly was placed at the end of the beamline in an in-air experimental section, with a separation of about \SI{50}{\centi\meter} from the beam exit window. A pair of quadrupole magnets is used to focus the beam in the experimental area. A Turbo-ICT with a resolution of $\sim1\%$ at \SI{100}{\pico\coulomb} allows the bunch charge to be measured right before the window~\cite{Lensch:2023wgx}. In addition to the data recorded with the EDS, data from the accelerator beam diagnostics were recorded through the main accelerator control system DOOCS~\cite{Hensler:1996ppq}. 

\begin{figure}[tbh]
    \centering
    \includegraphics[width=\linewidth]{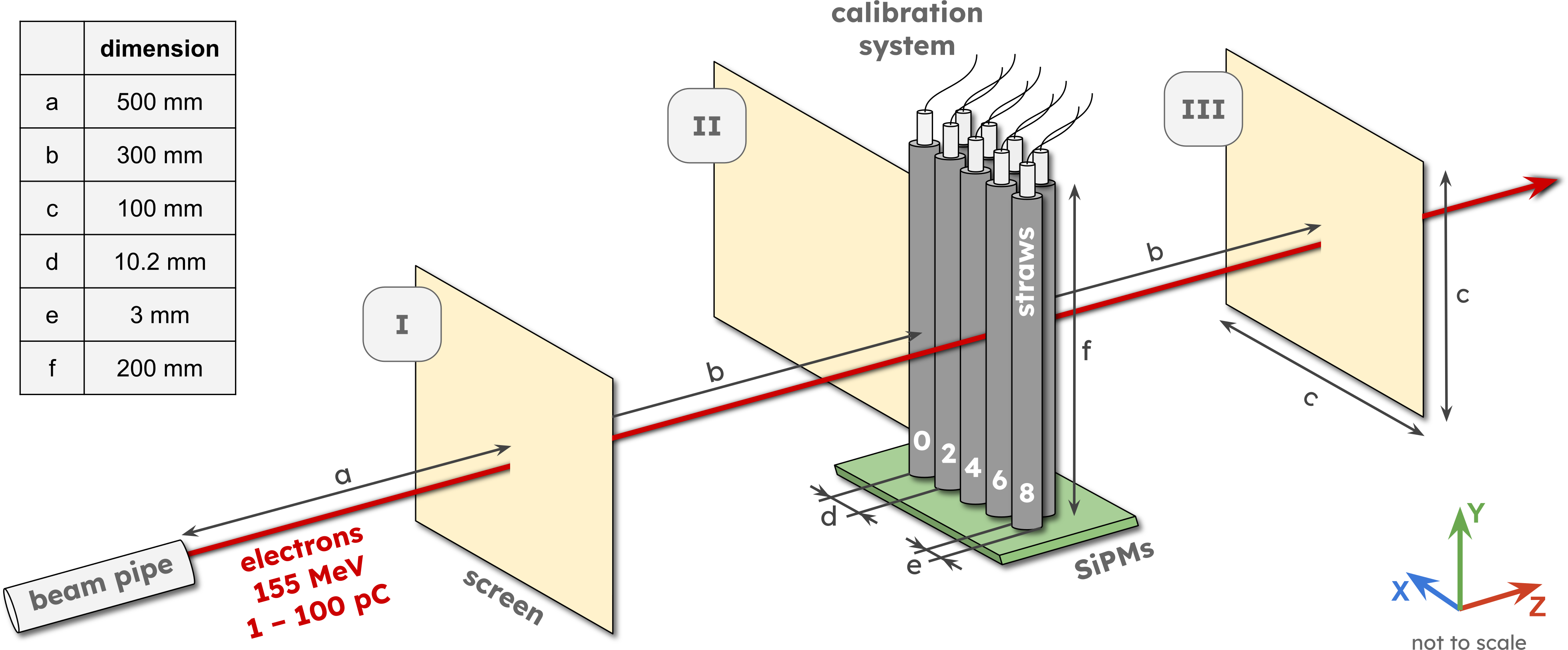}
    \caption{Sketch of the EDS setup at ARES. The setup consists of three identically sized scintillating screens placed at locations I, II, and III, a set of nine straws, a mechanical frame with one rotational and two linear stages, and an LED calibration system.}
    \label{fig:ares_sketch}
\end{figure}

As shown in figure~\ref{fig:ares_sketch}, three scintillating screens were placed within the prototype assembly, with one located \SI{30}{cm} in front of the straws (I), a second one located at the straw position (II), and a third one was located \SI{30}{cm} after the straws (III). The screens I and II were placed on linear stages and could be transversely moved out of the beam axis. The straws, composed of stainless steel with an inner diameter of \SI{3}{mm} and a wall thickness of \SI{0.1}{mm}, were placed perpendicularly to the beam axis. The entire straw frame was placed on the same linear stage as the screen at location II and could be rotated between \SI{0}{\degree} and \SI{30}{\degree} around the horizontal $x$-axis as indicated in figure~\ref{fig:straw_concept}. Each straw was equipped with an optical fiber for LED calibration.


Three different GadOx screen types from the DRZ product range by MCI Optonix~\cite{DRZ} were tested. Their properties, as provided by the manufacturer, are shown in table~\ref{tab:screens}. The different screen types provide a trade-off between brightness and spatial resolution, with thicker phosphor layers providing higher brightness but lower resolution.

\begin{table}[tbh]
\centering
    \begin{tabular}{c||c|c|c|c}
    DRZ type & \begin{tabular}[c]{@{}c@{}}phosphor Layer\\ thickness $[\mathrm{\upmu m}]$ \end{tabular} & \begin{tabular}[c]{@{}c@{}}phosphor layer\\ density $[\mathrm{mg/cm^2}]$\end{tabular} & \begin{tabular}[c]{@{}c@{}}relative\\ brightness\end{tabular} & \begin{tabular}[c]{@{}c@{}}relative \\ resolution
    \end{tabular}\\
    \hline \hline
    High & 310 & 145 & 158\% & 33\% \\
    \hline
    Plus & 208 & 100 & 120\% & 73\% \\
    \hline
    Standard & 140 & 68 & 100\% & 100\%
    \end{tabular}
    \caption{Properties of the different scintillating screen types for X-ray photons, as provided by the manufacturer~\cite{DRZ}. The resolution is based on the modulation transfer function at 2 line-pairs per millimeter~\cite{AHMED2015435}, with larger values corresponding to better spatial resolution. Relative values are given with respect to the Standard type.}
    \label{tab:screens}
\end{table}

At LUXE, it is essential that the scintillating screen produces sufficient light for the cameras to observe the full Compton spectrum, while the position resolution should be high enough to achieve a relative energy resolution of $<2\%$. Understanding the differences between the different screen types is therefore crucial to achieving optimal performance of the EDS. While the manufacturer specifications given in table~\ref{tab:screens} are based on X-ray measurements, the response of scintillating screens under electron irradiation may differ in terms of both brightness and resolution. DRZ scintillating screens have previously been tested with electron beams at energies of \SI{30}{\mega\electronvolt} and \SI{40}{\mega\electronvolt}~\cite{Schwinkendorf:2019zrz, Liu:2026nln}. 

At ARES, measurements were performed to confirm that the screen response is independent of the beam position on the screen. Additionally, the linearity of the screen response for increasing beam charge was also measured. From these measurements, the relative brightness of the different screen types for the ARES beam charges was extracted. The position resolution of the screen types was not measured.

Images of the scintillating screens were captured by a 2K-camera with a $f = \SI{50}{mm}$ lens and an optical filter, as described in section~\ref{sec:screen}. A single camera setting was used for all measurements. It was chosen such that the CMOS sensor did not saturate under any measurement conditions. Each of the different DRZ screen types was tested at all three screen locations.

To obtain a spatial calibration of the images, a scale attached to each scintillating screen was periodically imaged using an increased camera exposure time, making the scale markings visible. These reference images were used together with OpenCV~\cite{opencv_library} to perform a perspective transformation, allowing the beam spot to be viewed approximately head-on. After the transformation, the scale markings were used to determine the pixel-to-distance conversion factors, which yielded \SI{52.7(5)}{\micro\meter\per\pixel} and \SI{160(2)}{\micro\meter\per\pixel} for screen locations I and III, respectively.

The beam parameters were tuned to obtain a small horizontal width at the the straw detector, maximizing the relative charge passing through the straws. Since the beam profile is not exactly Gaussian, the beam width is characterized by the full width at half maximum (FWHM). Since screen I was also in the beam during measurements at screen location II, additional scattering prevented an independent determination of the beam width. Because the DRZ screen type did not significantly affect the measured beam size, the measurements of the different screen types were combined separately at each screen location using a weighted mean of the FWHMs determined for the individual measurements. The average horizontal widths at screen locations I and III are \SI{1.1(1)}{\milli\meter} and \SI{4.7(2)}{\milli\meter}, respectively. To estimate the beam width in the plane of the straw detector, corresponding to location II, a linear interpolation was applied between the measurements at the screen locations I and III. This procedure yielded a horizontal FWHM of \SI{2.9(1)}{\milli\meter}.

\begin{figure}[tbh]
    \centering
    \includegraphics[width=0.6\linewidth]{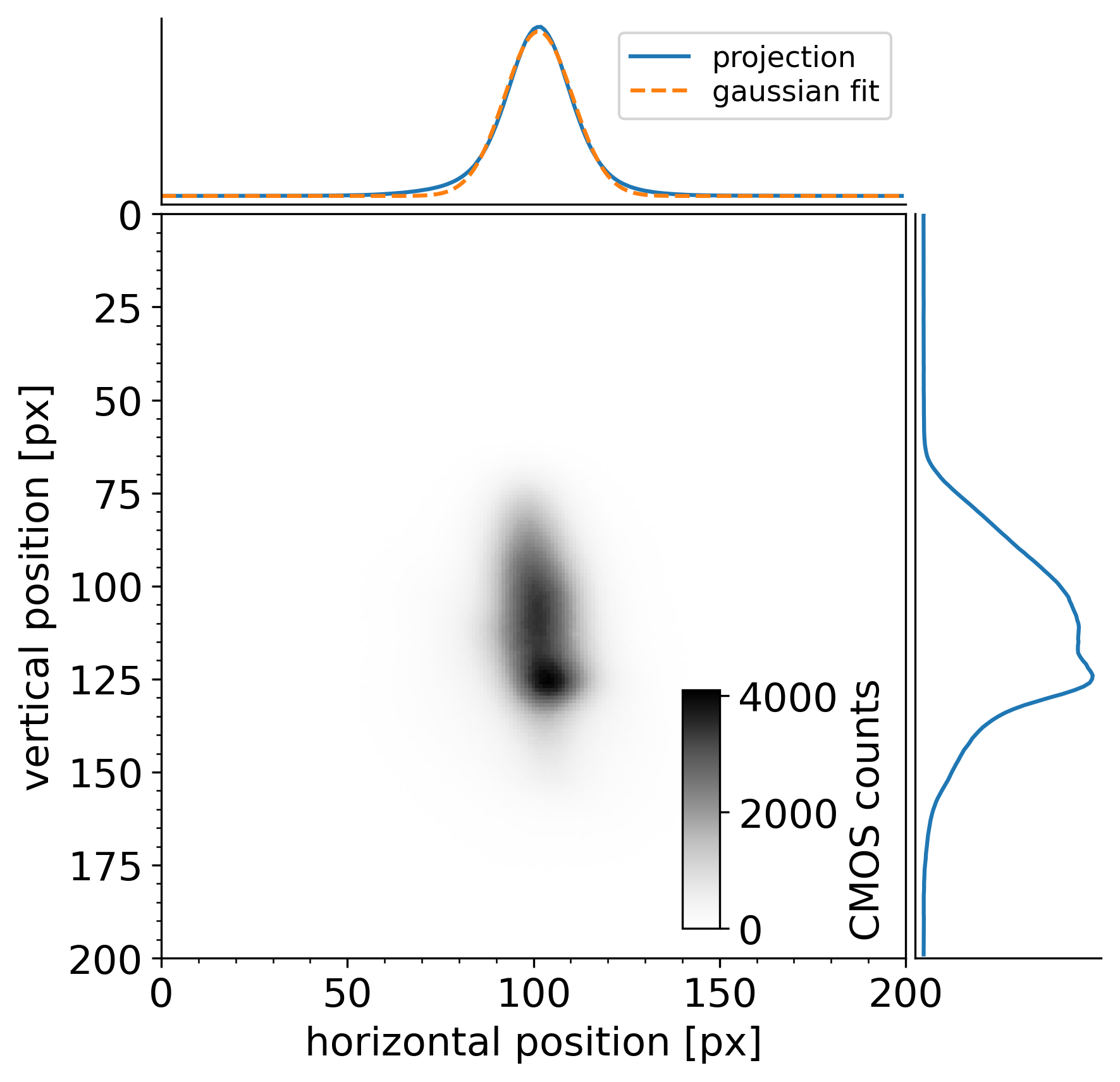}
    \caption{A single-shot image of screen location I taken at ARES showing the ROI with the beam spot in the center (black). Also shown are the horizontal and vertical beam profiles, together with a Gaussian fit to the horizontal profile.}
    \label{fig:ares_screen}
\end{figure}

The following scintillating screen measurements were all performed at screen location I. The response of the different scintillating screen types was compared using the recorded CMOS counts, which are proportional to the number of photons detected by the CMOS sensor. A region of interest (ROI) in the images was chosen, identical for every image, encompassing the beam spot. An example is shown in figure~\ref{fig:ares_screen}. The CMOS counts in this region were summed to obtain the light yield of the screen, called the signal integral. 

\begin{figure}[tbh]
    \centering
    \includegraphics[width=0.7\linewidth]{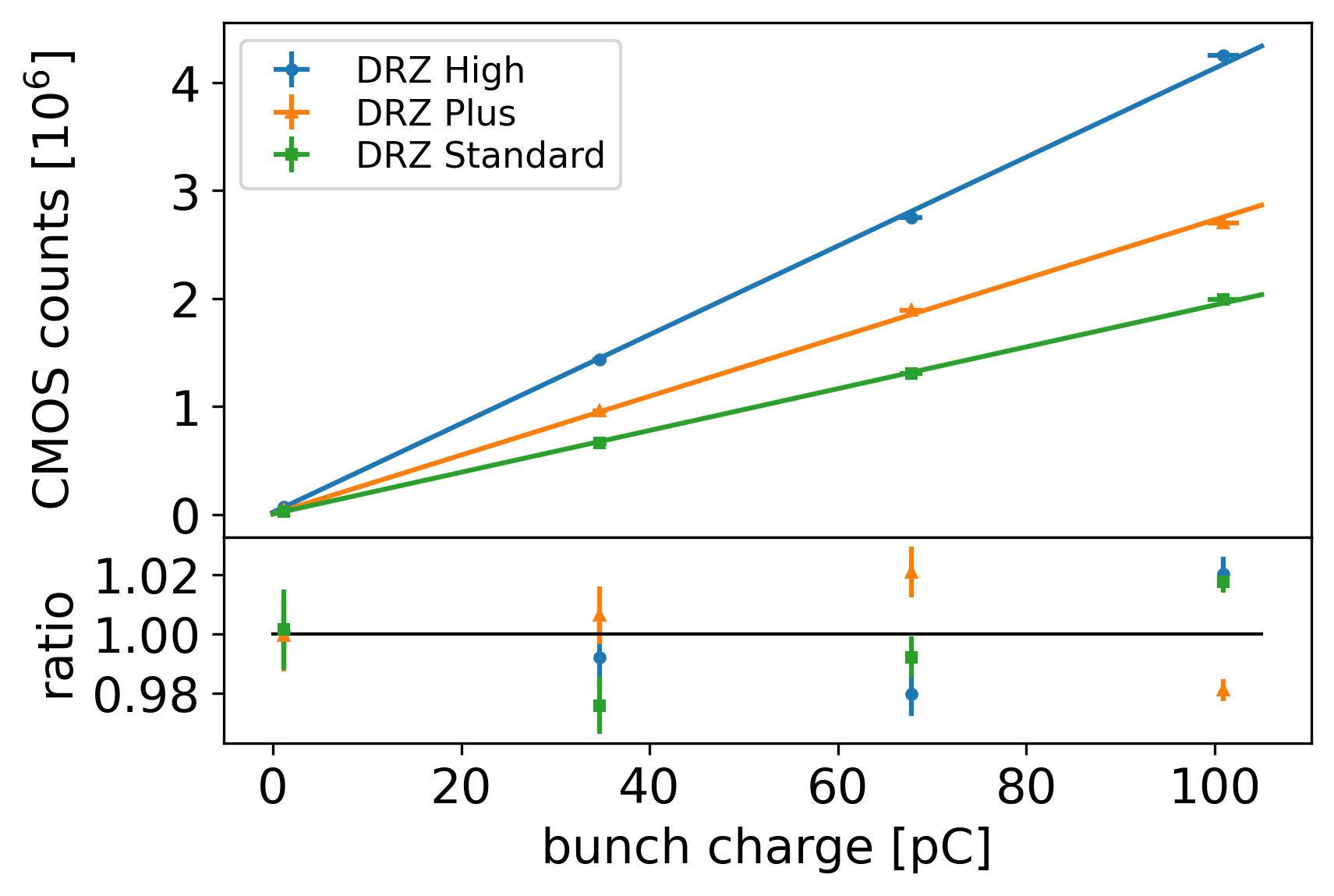}
    \caption{A comparison of the light emission for different beam charges and screen types, as measured by the camera. A linear fit of the data points is also included. The ratio plot shows the ratio between the data points and the linear fits for each scintillating screen.}
    \label{fig:screencompar}
\end{figure}

In figure~\ref{fig:screencompar}, a measure of the linearity of the screen response is shown. The signal integral for each screen type and charge combination is shown together with a linear fit applied to the data. The fits were not constrained to pass through the origin and yielded small positive offsets, corresponding to less than 0.6\% of the signal at \SI{100}{\pico\coulomb}. These offsets are attributed to background contributions to the measured signal and can be accounted for during detector calibration. The ratio between the data and the fitted line shows no deviations from a linear response larger than 3\%. The uncertainty of the beam charge measured by the beamline’s Turbo-ICT contributes to the uncertainty of the fitted calibration slope. Including the uncertainties in both coordinates, the fitted calibration slopes have relative uncertainties of approximately 1.3\% for all three screen types, below the $2.5\%$ uncertainty targeted by LUXE. 

The relative brightness and the CMOS calibration for this setup were obtained from these measurements and are shown in table~\ref{tab:screen_results}. The relative brightness was calculated by normalizing each DRZ High and DRZ Plus data point to DRZ Standard at equivalent bunch charges, then computing the weighted mean for each DRZ type. The charge uncertainties were not included in this estimate. It can be seen that the relative brightness measured with electrons at ARES differs significantly from those provided by the manufacturer, with DRZ High being over twice as bright as DRZ Standard compared to only 1.58 times for X-rays (see table \ref{tab:screens}). The CMOS calibration factor was determined from the inverse slope of the linear fit between the integrated CMOS signal and the bunch charge. It corresponds to the number of incident electrons required to produce one CMOS count. The calibration factor is specific to the camera configuration used for the measurements.

\begin{table}[tbh]
    \centering
    \begin{tabular}{c||c|c}
    DRZ type & \begin{tabular}[c]{@{}c@{}}relative\\ brightness \end{tabular} & \begin{tabular}[c]{@{}c@{}}CMOS\\ calibration \end{tabular}\\
    \hline \hline
    High & $(216 \pm 1)\%$ & $(144\pm2)~\var{N}{e}$\\
    \hline
    Plus & $(138 \pm 1)\%$ & $(224\pm2)~\var{N}{e}$\\
    \hline
    Standard & $100\%$ & $(316\pm3)~\var{N}{e}$
    \end{tabular}
    \caption{Properties of the different screen types measured at ARES. }
    \label{tab:screen_results}
\end{table}

\begin{figure}[tbh]
    \centering
    \includegraphics[width=0.7\linewidth]{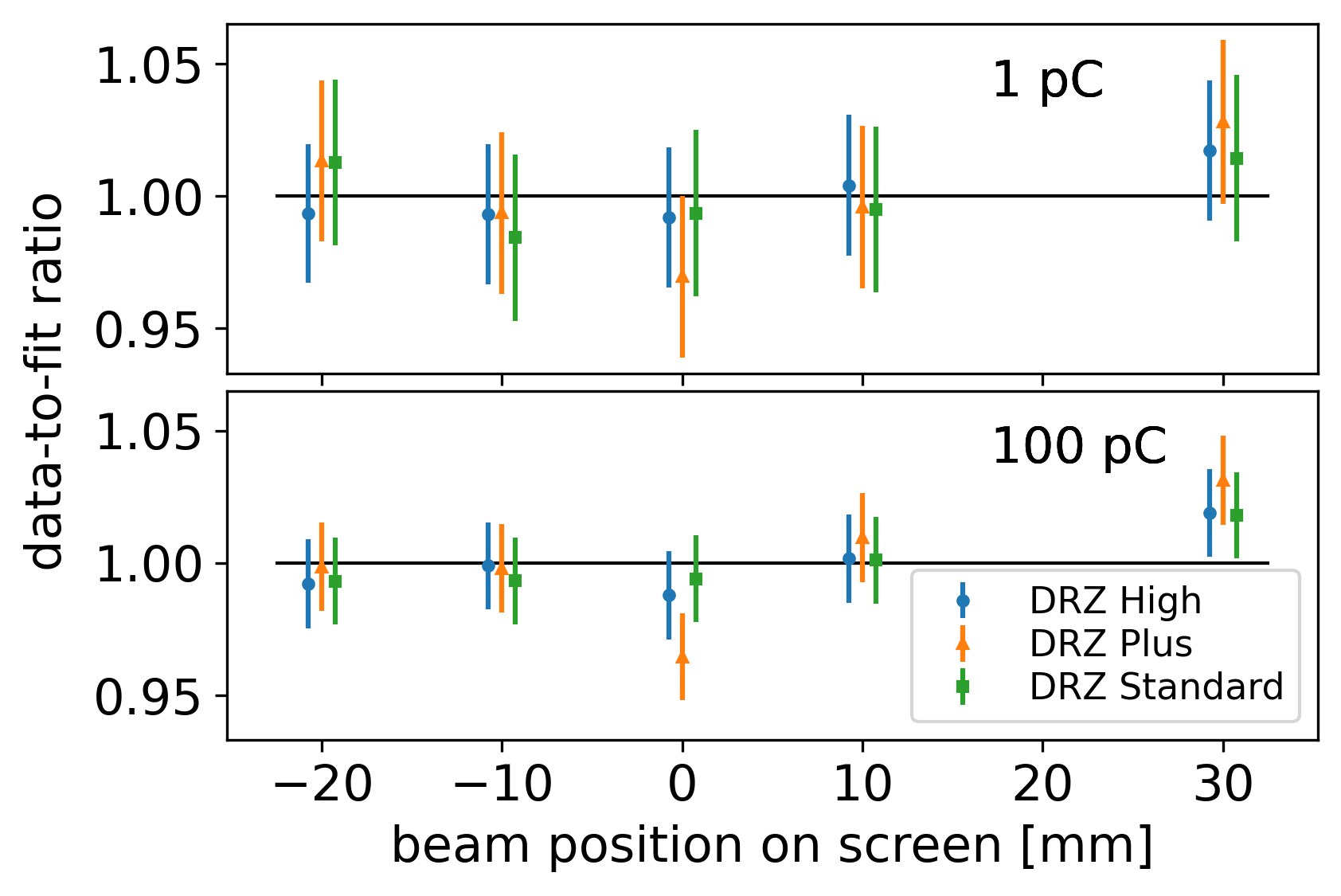}
    \caption{Measured uniformity of the screen response for different beam positions. The data points were obtained by dividing each measured value by the fitted constant value per screen type. Only the measurements for the beam charges \SI{1}{pC} and \SI{100}{pC} are shown here. The data points for the DRZ High and DRZ Standard are slightly offset from their nominal $x$-position to improve the readability of the plot.}
    \label{fig:posscan}
\end{figure}

Measurements of the uniformity of the screen response were performed by varying the horizontal position of the screen at location I with respect to the beam using the linear stage and measuring the signal integral. Figure \ref{fig:posscan} shows the ratio of the signal integrals for different positions and the fitted constant values for the data points of each of the different screen types, with each data point considering $1000$ images. The signal integral is independent of the beam position to within 3\% for all tested beam charges. Similar variations with beam position are observed at different bunch charges, suggesting a small systematic component, although the deviations are compatible with the measurement uncertainties. For clarity, only the measurements for beam charges of \SI{1}{pC} and \SI{100}{pC} are shown.

The EDS at LUXE is aligned such that the primary electron beam bypasses the scintillating screen, while only electrons scattered in the laser interaction are intercepted. Simulations predict a peak charge density of up to \SI{100}{\pico\coulomb\per\milli\meter\squared} on the screen during normal operation. The charge densities reached during the ARES measurements were below this expected maximum. Previous studies have demonstrated that GadOx screens remain linear beyond \SI{350}{\pico\coulomb\per\milli\meter\squared}~\cite{Keeble:2019eqo, Bauche:2019vjt}, indicating that the expected charge densities at LUXE are well within the linear response regime.

The second detector technology investigated for the EDS is the straw-based Cherenkov detector. The prototype used at ARES comprises nine air-filled stainless-steel straws, which are placed in a two-row configuration with \SI{10.2}{\milli\meter} spacing. The two-row arrangement partially covers the gaps between adjacent straws, reducing the uncovered gaps in the beam view to \SI{1.9}{\milli\meter}. The five straws with even numbers are in the front row, whereas the four straws with odd numbers are in the rear row. 

Two variants of the same SiPM model were tested. The Hamamatsu MPPC S14160-30xx is a SiPM with an active area of $3 \times \SI{3}{mm^2}$, a spectral range from about $\lambda_1 = \SI{300}{nm}$ to $\lambda_2 = \SI{900}{nm}$ and a gain of the order of $10^5$. The two variants have pixel pitches of \SI{10}{\micro\meter} ("SiPM-10") and \SI{15}{\micro\meter} ("SiPM-15")~\cite{hamamatsu_photonics_kk_hamamatsu_2023}. Compared to SiPM-15, SiPM-10 contains approximately twice as many pixels. This increases the saturation threshold, since more pixels are available to detect individual photons, but results in a lower gain due to the smaller pixel size. The SiPM-10 sensors were attached to straws 0, 1, 2, and 7, and the SiPM-15 were attached to straws 3, 4, 5, 6, and 8. In this configuration, each type of SiPM had at least two straws in each row. 

\begin{figure}[tbh]
    \centering
    \includegraphics[width=0.9\linewidth]{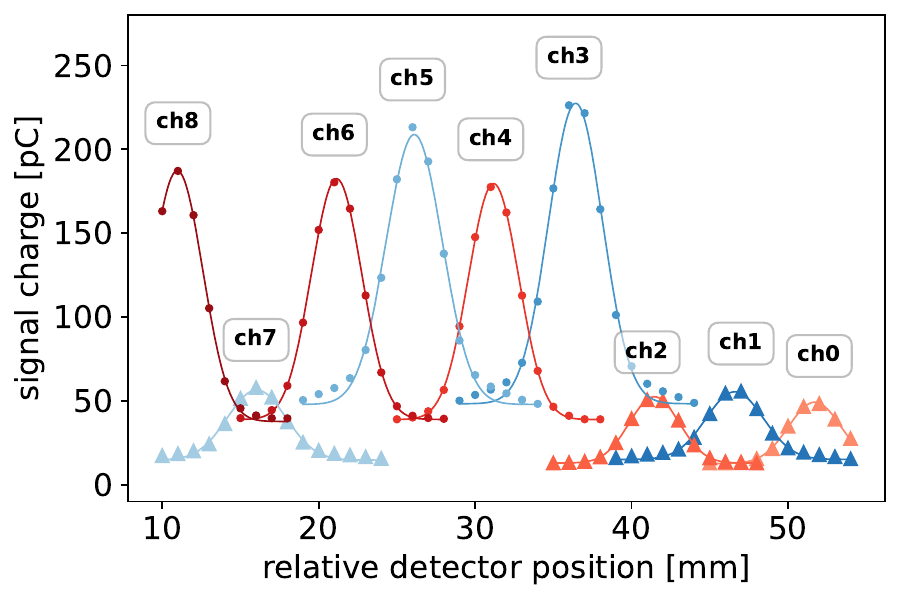}
    \caption{Scan of the nine straws through the Gaussian beam with a bunch charge of \SI{110}{\pico\coulomb}. The channels 0, 1, 2, and 7 are equipped with SiPM-10 (\ding{115}), and channels 3, 4, 5, 6, and 8 with SiPM-15 (\ding{108}). The even channel numbers represent front-row straws (red), and the odd channel numbers represent rear-row straws (blue). The solid lines correspond to least-squares fits using equation~\eqref{eq:l_total}. }
    \label{fig:yscan_ares}
\end{figure}

To characterize the signal response of the detector, it was moved transversely through the beam and its profile was scanned dynamically as shown in figure~\ref{fig:yscan_ares}. The comparison between the two straw rows and the two SiPM types is based on measurements at all investigated bunch charges. Since no systematic dependence on bunch charge was observed, the average ratios across all charge settings are summarized in table~\ref{tab:straw_comparison}. The straws in the rear row measured a larger signal, corresponding to a higher number of charged particles traversing the straws. This is likely caused by secondary particles produced when the beam passes through the front-row straws. The two sensor types differ significantly in signal amplitude. This is expected since both sensor types were operated with the same bias voltage rather than the same gain. 

\begin{table}[tbh]
    \centering
    \begin{minipage}{0.47\linewidth}
        \centering
        \begin{tabular}{l|c|c}
             & SiPM-10 & SiPM-15 \\
            \hline
            rear / front & 1.07 & 1.16
        \end{tabular}
    \end{minipage}
    \hfill
    \begin{minipage}{0.47\linewidth}
        \centering
        \begin{tabular}{l|c|c}
             & Front & Rear \\
            \hline
            SiPM-15 / SiPM-10 & 3.9 & 4.2
        \end{tabular}
    \end{minipage}
    \caption{Average ratios of the baseline-subtracted signal amplitude between the two straw rows and both SiPM types across all investigated bunch charges.}
    \label{tab:straw_comparison}
\end{table}

The baseline offset visible in the scan indicates that the photosensors also measured some signal when the beam was not passing through a given straw. The baseline was found to increase approximately linearly with the bunch charge, indicating the presence of a beam-induced background. This background is likely caused by beam halo or electrons scattered at the beam-exit window, which may also directly hit the SiPMs and produce a signal. Beam-induced electronic signals provide another possible contribution. This background was further investigated during the FACET-II measurements described in section~\ref{sec:facet}.

The transverse position scan of the detector through the beam also allows the beam size to be determined, taking into account the round straw geometry by using equation~\eqref{eq:l_total}. An average of all charge settings and straws yields horizontal FWHMs of \SI{3.27(7)}{\milli\meter} and \SI{3.78(8)}{\milli\meter} for the front and rear rows, respectively. The quoted uncertainties include the scaling of the statistical fit uncertainties according to the observed spread of the measurements. In addition to the larger signal amplitude discussed above, the rear row also yields a larger reconstructed beam width, which is consistent with beam broadening and scattering induced by the front-row straws. Based on these reconstructed beam widths, the fractions of the bunch charge traversing a centered straw were determined by integrating the Gaussian beam profile within the straw radius of \SI{1.5}{\milli\meter}, yielding \num{0.72(1)} and \num{0.65(1)} for the front and rear rows, respectively, with uncertainties propagated from the reconstructed beam widths. These factors were applied to the charge measurements. The width measured by the front-row straws is larger than the \SI{2.9(1)}{\milli\meter} obtained by interpolation between the two scintillating screen measurements. A similar systematic difference between the two detectors was observed during the FACET-II measurements and is discussed in section~\ref{sec:facet}.

\begin{figure}[tbh]
    \centering
    \begin{subfigure}[t]{0.48\textwidth}
        \centering
        \includegraphics[width=\linewidth]{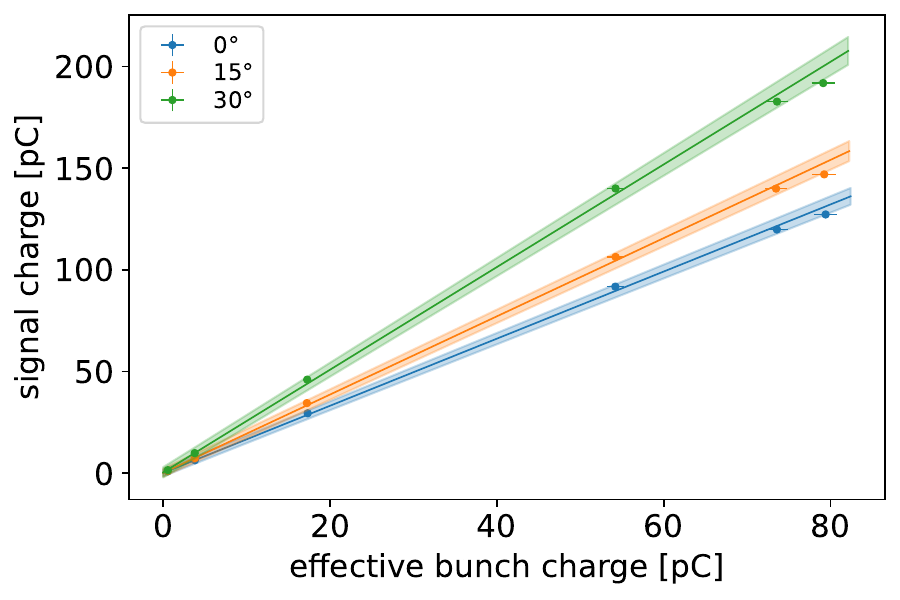}
        \caption{~}
        \label{fig:straw_angle_charge}
    \end{subfigure}
    \hfill
    \begin{subfigure}[t]{0.48\textwidth}
        \centering
        \includegraphics[width=\linewidth]{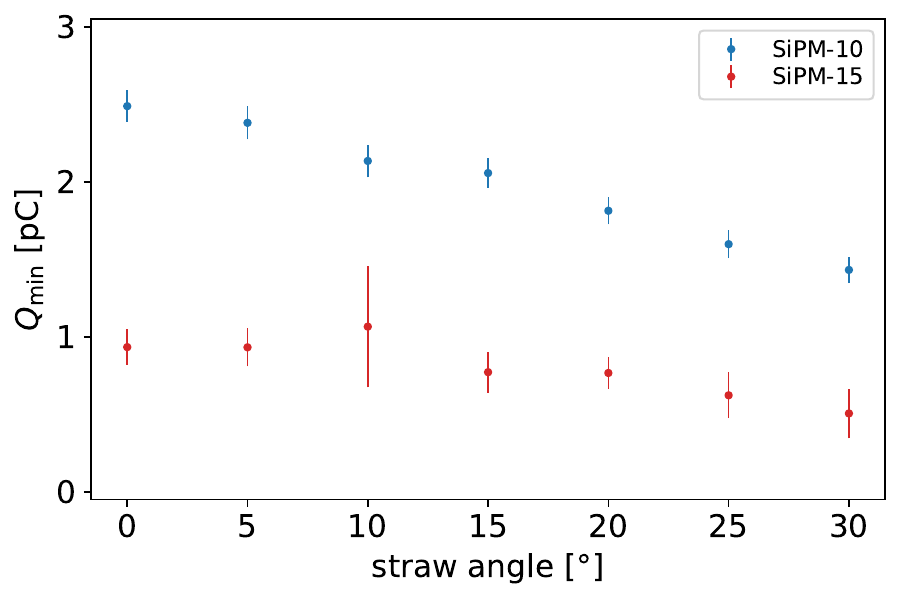}
        \caption{~}
        \label{fig:q_min}
    \end{subfigure}
    \caption{Results from the charge scan measurements at ARES. a) Signal amplitude of straw channel four, equipped with SiPM-15, for different effective bunch charges and straw angles. The solid lines correspond to a linear fit with its uncertainty band. b) $\var{Q}{min}$ as a function of straw angle for SiPM-10 (channel 2, blue) and SiPM-15 (channel 4, red).}
    \label{fig:ares_straw_charge_scan_plots}
\end{figure}

The response of the straw detector for one channel of each type of SiPM was measured as a function of the beam charge for different straw angles, as shown in figure~\ref{fig:straw_angle_charge}. The signal intensity increases linearly with the beam charge. The slight leveling off of the signal response at high bunch charges indicates the onset of SiPM saturation. Increasing the straw tilt from \SI{0}{\degree} to \SI{30}{\degree} increases the signal by 53\%. 

The minimum detectable bunch charge $\var{Q}{min}$ of the straw detector was determined as the charge at which a signal could be measured with $5\sigma$ confidence above background. The background level and its single-shot standard deviation were determined from measurements without an electron beam, and the corresponding $5\sigma$ signal threshold was converted to bunch charge using the linear charge calibration. The results are shown in figure~\ref{fig:q_min}. The minimum detectable charge is generally lower for the more sensitive SiPM-15 photosensor and decreases with increasing straw angle, as more Cherenkov photons reach the SiPM. The lowest value obtained in this measurement was $\var{Q}{min} = \SI{0.5(2)}{\pico\coulomb}$ for SiPM-15 at a straw angle of \SI{30}{\degree}. This corresponds to approximately \num{3e6} electrons, while the LUXE detector is required to be sensitive to electron fluxes as low as \num{e4} electrons. To achieve the additional sensitivity required in detector regions with low expected electron fluxes, a Cherenkov medium with a higher photon yield can be selected. This approach was investigated during the subsequent FACET-II measurement campaign using solid glass rods, described in section~\ref{sec:facet}. Furthermore, the sensitivity could be further optimized by adapting the SiPM gain and pixel pitch, as well as the angle of the Cherenkov medium with respect to the beam, to the expected electron flux in different detector regions.

\subsection{FACET-II}\label{sec:facet}

The latest prototype was tested at FACET-II at SLAC, in collaboration with the E-320 strong-field QED experiment. While the ARES campaign characterized individual detector components using a low-energy, monoenergetic electron beam, the FACET-II campaign enabled the combined EDS to be tested at high beam energy using an electron spectrum dispersed by a magnetic spectrometer. The beam conditions at FACET-II are summarized in table~\ref{tab:facilityParameters}. The key components for our tests in the beamline are an interaction chamber followed by a magnetic imaging spectrometer consisting of three quadrupole magnets and a dipole magnet. The E-320 experiment collides the FACET-II electron beam inside the interaction chamber with a laser pulse that can provide a peak power of up to \SI{10}{\tera\watt}, a pulse duration of about \SI{42}{\femto\second} FWHM, and a waist at the interaction point of \SI{2}{\micro\meter}. At FACET-II, the scintillating screen and Cherenkov detector performance, as well as the effect of radiation on the SiPMs, were tested using the electron primary beam. Data from the accelerator beam diagnostics were recorded through the EPICS main accelerator control system~\cite{epicscontrols_experimental_2024, Gessner:2021wgi}. During the measurement campaign, the electron energy of \SI{9.77(8)}{\giga\electronvolt} was measured with the energy spectrometer and a bunch charge of \SI{1.47(2)}{\nano\coulomb} was measured with a toroid beam charge transformer. 

\begin{figure}[tbh]
    \centering
    \includegraphics[width=\linewidth]{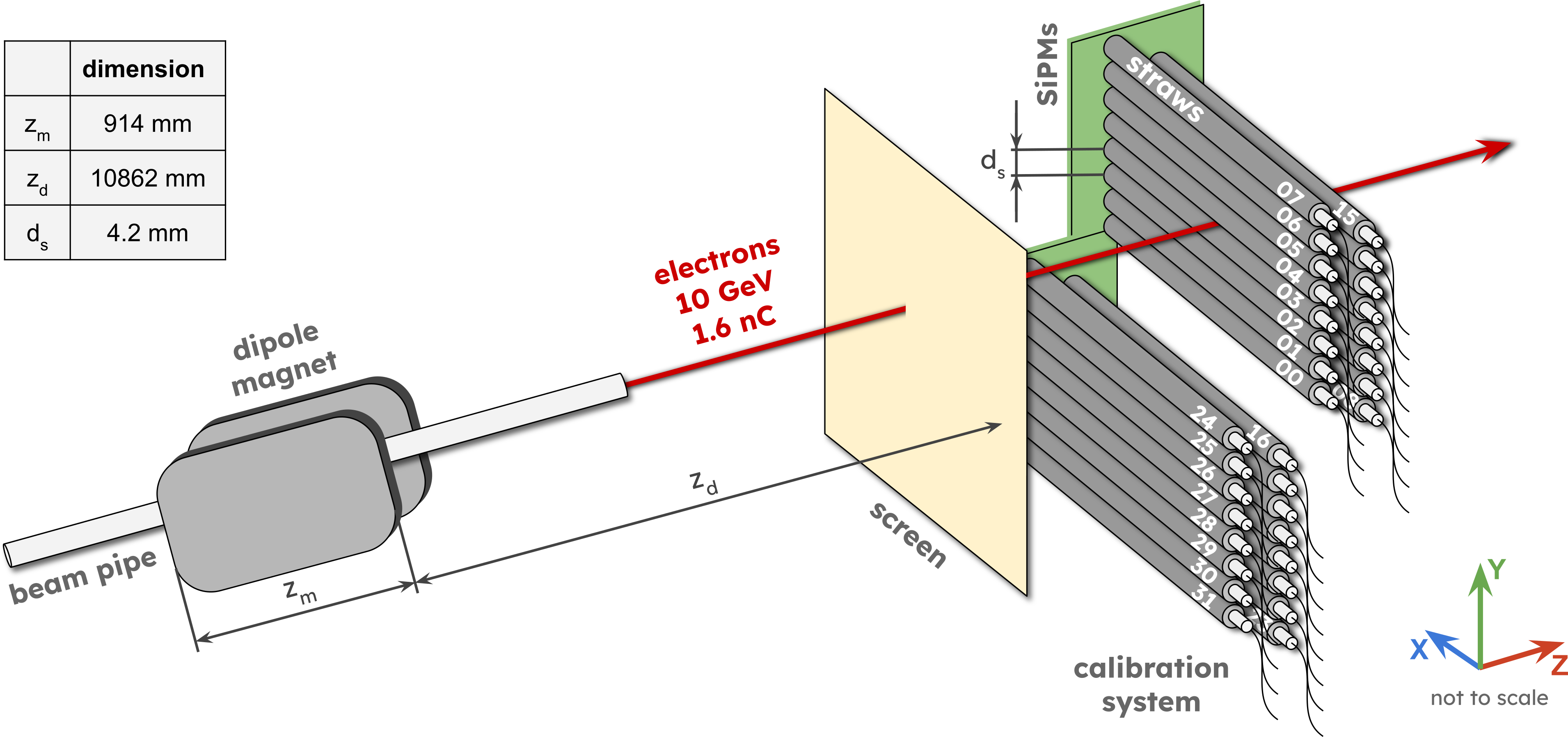}
    \caption{Sketch of the EDS setup at FACET-II. The electrons pass a spectrometer dipole magnet and are deflected downwards and the energy spectrum is therefore expected to be measured vertically. The EDS prototype is placed in the in-air experimental area and comprises a scintillating screen, a stainless-steel-straw, and a glass-rod module. }
    \label{fig:luxe320_sketch}
\end{figure}

The detector designed to fit the FACET-II facility is shown in figure~\ref{fig:luxe320_sketch}. After the electrons are accelerated to their final energy, they enter an in-vacuum interaction chamber where the electron--laser interaction point is located. It is followed by a \SI{0.2}{T} spectrometer dipole magnet that deflects the electrons vertically with respect to their energy. About \SI{11}{m} after the dipole magnet, the EDS is placed in the in-air experimental area before the main beam dump. The electrons pass a scintillating screen that is attached in front of two sets of straw detector modules. As shown in figure~\ref{fig:edsPhotoFacet}, the scintillating screen and straw setup is placed on linear and rotational stages. These stages allow for linear movements in the horizontal and vertical direction, as well as rotations around the vertical axis to change the straw angle with respect to the electron beam. 

The scintillating screen setup at FACET-II includes a 2K-camera with an $f=\SI{50}{mm}$ lens and an optical filter. The camera is mounted \SI{1.2}{m} upstream of the screen. The setup contains one $\SI{10}{cm} \times \SI{10}{cm}$ DRZ Standard screen. The choice of the screen type with the lowest light yield was motivated by the higher bunch charge at FACET-II compared to ARES. On the outer rim, a \SI{1}{\centi\meter} wide scale is placed, which reduces the effective size of the scintillating screen to $\SI{8}{\centi\meter} \times \SI{8}{\centi\meter}$. The scale enables correction for the perspective view of the camera and calibration of the position and size of the beam profile and Compton spectrum, regardless of the vertical position of the screen-straw setup. 

The straw detector is composed of two modules of densely arranged straws, as shown in figure~\ref{fig:luxe320_sketch}. Each module comprises two tiers of eight straws, with a center-to-center distance of $d_s = \SI{4.2}{\milli\meter}$. This configuration leaves a \SI{1}{\milli\meter} gap between adjacent straws within each tier. These gaps are covered by the straws in the second tier, which is transversely offset by half the straw spacing. For the FACET-II spectrometer geometry, the \SI{3}{\milli\meter} straw diameter corresponds to a single-straw energy resolution of approximately 4\% at the \SI{8}{\giga\electronvolt} linear Compton edge. To maximize the sensitivity in the low-energy tail of the Compton spectrum, where the electron flux and consequently the total Cherenkov signal are lowest, two complementary approaches were implemented. The straw modules on the high-energy side of the spectrum accommodate steel straws identical to those used at ARES, while those on the low-energy side consist of glass rods as described in section~\ref{sec:eds}, providing a higher Cherenkov light yield. In addition, all tests at FACET-II were performed using SiPM-15 with a \SI{15}{\micro\meter} pixel pitch (see section~\ref{sec:ares}). Their higher intrinsic gain allows operation at a lower bias voltage while maintaining sufficient signal amplitude.

An energy calibration was performed by measuring the beam position on the screen for different dipole strengths. The deflection of an electron from its original axis due to a dipole magnet can be approximated as
\begin{align}\label{eq:dipole_deflection}
    y_d = \left(\frac{z_m}{2}+z_d\right) \, z_m c \, \frac{e B }{E} \, ,
\end{align}
where $e$ is the elementary charge, $B$ the dipole field strength, and $E$ the electron energy. The length of the dipole field $z_m$ and the drift distance to the detector $z_d$ are obtained from the FACET-II geometry. 

\begin{figure}[t]
    \centering
    \includegraphics[width=0.8\linewidth]{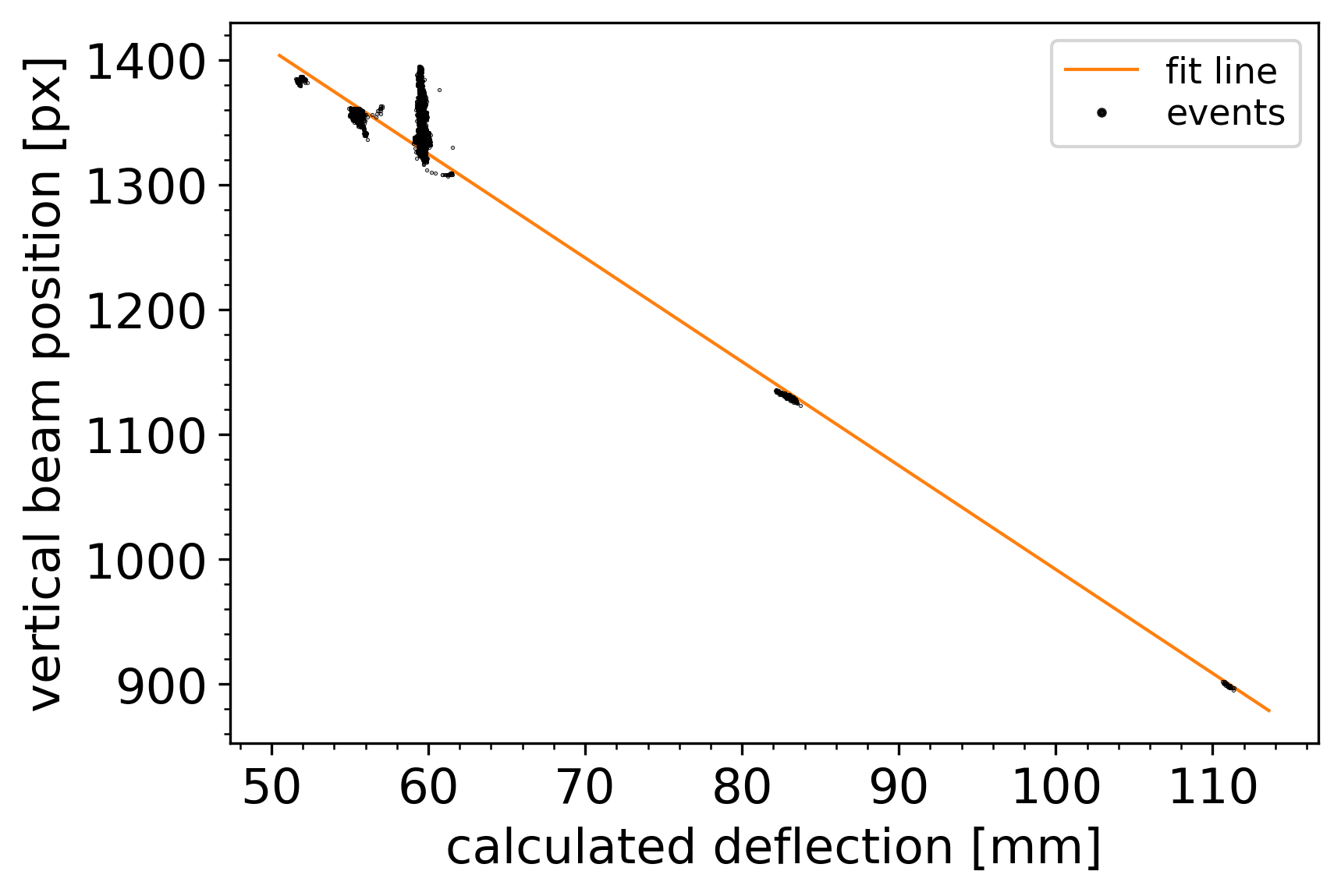}
    \caption{Vertical beam position measured with the scintillating screen detector as a function of the calculated beam deflection for the five dipole settings. The solid line shows the orthogonal distance regression fit using the individual uncertainties of all measurements. For clarity, individual error bars are omitted. The typical measurement uncertainties are \SI{90}{\micro\meter} for the calculated deflection and \SI{37}{\pixel} for the vertical beam position. }
    \label{fig:dipole_scan}
\end{figure}

During the measurements with different dipole strengths, the energy of the electron beam was also obtained from the FACET-II beamline diagonstics. This allowed for the prediction of the expected $y_d$ at the position of the EDS screen. The calculated deflection against the measured vertical beam position in pixels on the screen is shown in figure~\ref{fig:dipole_scan}. The majority of events (73\%) were recorded at the nominal dipole setting, corresponding to the event cluster at $\sim \SI{60}{mm}$. The spread in the calculated beam deflection at a given dipole setting originates primarily from variations in the measured beam energy, while the larger spread in the measured beam position is attributed to beam-position drifts. The measured beam position was fitted as a linear function of the calculated deflection, yielding a pixel-to-distance conversion factor of \SI{120.2(2)}{\micro\meter\per\pixel}. The conversion factor was also measured independently using the scale on the screen, resulting in a value of \SI{121(1)}{\micro\meter\per\pixel}, which is in good agreement. The energy axis can then be obtained from equation~\eqref{eq:dipole_deflection} using the measured pixel-to-distance conversion factor. 

The position resolution of the 2K-camera at FACET-II (\SI{120}{\upmu m/px}) is similar to that expected at LUXE (\SI{136}{\upmu m/px} \cite{LUXE:2023crk}). It was only possible to resolve the Compton spectrum down to \SI{7.0}{GeV}, while the full spectrum is expected to extend further. To resolve this energy, it was necessary to change the camera settings to maximize the gain and increase the exposure time to \SI{50}{\milli\second} (the highest value where it was still possible to take images at \SI{10}{Hz}). Since LUXE will also operate in the range of laser intensities explored at E-320, similar rates of electron--laser interactions are expected. However, compared to FACET-II, LUXE will operate with a lower beam charge, while the Compton spectrum will be distributed over a larger spatial extent. Consequently, the signal detected per camera pixel is expected to be lower. Several approaches can be considered to compensate for this reduction in signal. These include the use of scintillating screens with a higher light yield, including alternative scintillator materials, as well as optimizing the optical readout, for example by adjusting the camera settings or the distance between the camera and the screen. The suitability of these approaches must be evaluated while preserving the required position resolution of the EDS. Furthermore, combining scintillating screens with different light yields and camera systems with different characteristics provides a means to optimize the sensitivity across the full Compton spectrum while maintaining the required position resolution. In such a configuration, particular care must be taken to minimize discontinuities in the reconstructed spectrum at the transition between the different detector regions.

\begin{figure}
    \centering
    \includegraphics[width=0.75\linewidth]{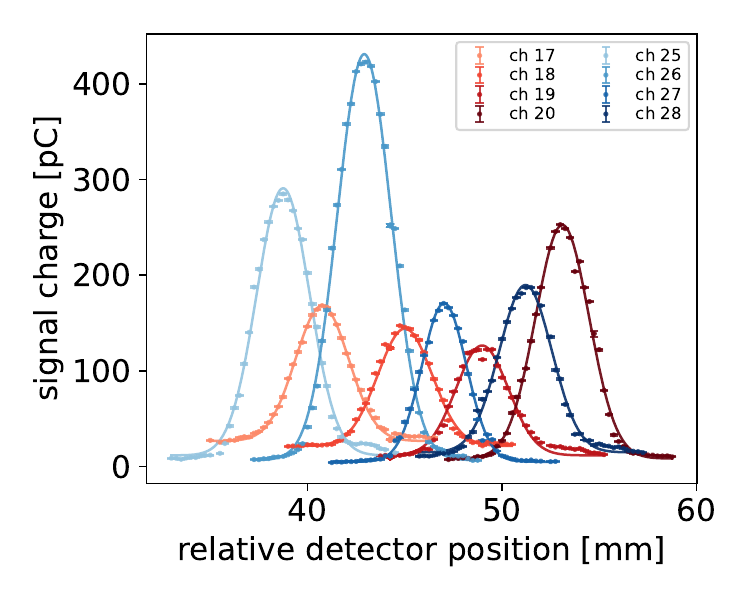}
    \caption{Linear detector position scan of a set of glass rods through a Gaussian beam. The graphs in blue represent the front row and the ones in red the rear row. The fit was performed using equation~\eqref{eq:l_total}. Due to accelerator instabilities during the measurement, systematic beam spot shifts might be larger than estimated in the displayed position uncertainty.}
    \label{fig:yscan_facet}
\end{figure}

When the detector was moved vertically through the beam, dynamic measurements of the vertical beam profile and a straw channel calibration were performed. Figure~\ref{fig:yscan_facet} shows the signal response of a position scan for a selection of glass rods. In analogy to the ARES analysis, equation~\eqref{eq:l_total} was used for the least-squares fit to account for the round straw geometry and the beam overlap. The mean FWHM of the beam was \SI{2.98(5)}{\milli\meter} and \SI{3.34(3)}{\milli\meter} for the steel straws and glass rods, respectively, with no significant difference between the front and rear rows. During the same measurement sequence, the beam spot size was also analyzed using the scintillating screen measurements, resulting in an average FWHM of \SI{2.3(1)}{\milli\meter}. 

As in the ARES measurement campaign, the beam width reconstructed with the straw detector is systematically larger than that obtained with the scintillating screen. The discrepancy was more pronounced at FACET-II, where larger shot-to-shot fluctuations of the beam parameters, particularly the beam position, were observed. These fluctuations are also visible as irregularities in the data of figure~\ref{fig:yscan_facet}. Since the straw profile is reconstructed from a sequence of measurements taken at different detector positions, such fluctuations can broaden the reconstructed profile. Systematic effects related to the straw-response model may also contribute, as indicated by the different widths reconstructed with the steel straws and glass rods. The origin of the discrepancy could not be conclusively identified and should be investigated in future beam-test campaigns and more detailed simulations. 

During the dynamic measurements, the SiPMs coupled to the glass rods were operated at a lower gain than those of the steel straws, as the higher Cherenkov light yield of the rods was expected to increase the signal intensity by a factor of about \num{e3}. However, the observed signal increase was significantly smaller than expected. This is likely due to the lower reflectivity of the glass rod wrapping compared to the steel straws, as well as the air gap between the glass rods and the SiPMs, which reduces the optical coupling efficiency. Since the glass-rod module was optimized for the low electron flux in the low-energy tail of the Compton spectrum, measurements with the primary beam required operation below the recommended bias voltage range of the SiPMs to avoid electronic saturation. Consequently, the expected increase in signal could not be quantified reliably. A dedicated characterization of the glass rods, analogous to that performed for the steel straws at ARES, is therefore required in future beam-test campaigns to quantify their sensitivity and optimize their implementation for LUXE.

\begin{figure}
    \centering
    \includegraphics[scale=1]{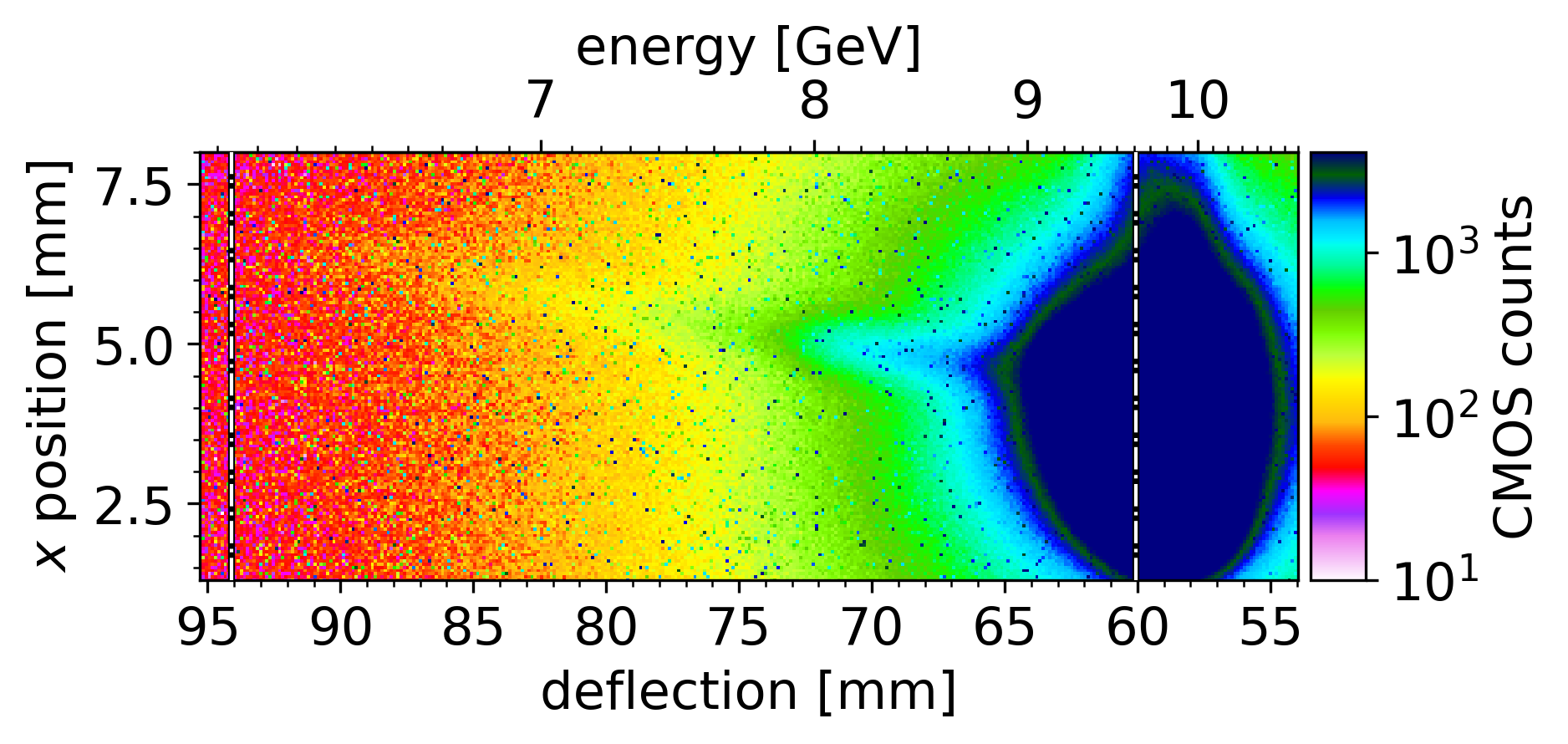}
    \caption{An image of the scintillating screen taken by the 2K camera at FACET-II. The primary beam spot is located on the right at 10 GeV, with a linear Compton spectrum emanating to the left to lower energies. The vertical dashed lines indicate the approximate boundaries of the glass-rod module of the Cherenkov detector. }
    \label{fig:screen_facet}
\end{figure}

\begin{figure}[tbh]
    \centering
    \includegraphics[width=0.8\linewidth]{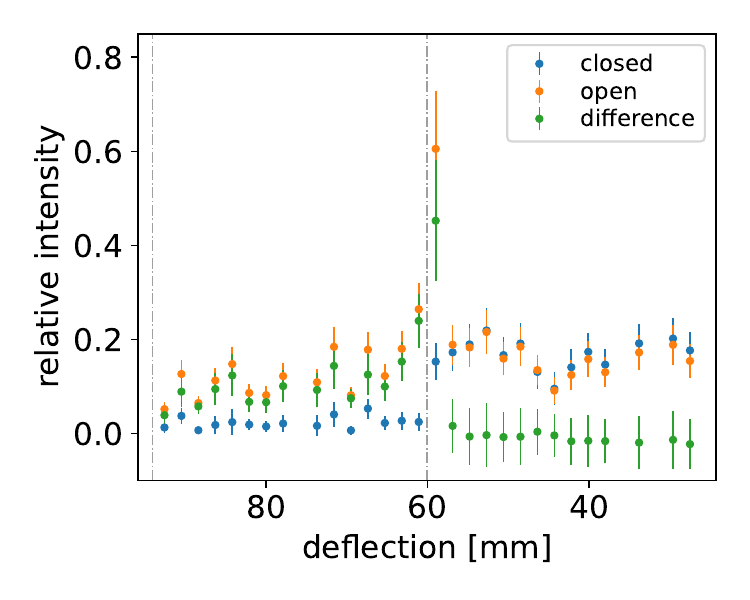}
    \caption{Comparison of the relative responses measured with the shutter closed (blue) and open (orange), together with their difference (green), for all channels. The dashed lines indicate the boundaries of the glass-rod module, with the steel straws located to its right.}
    \label{fig:facet_darkening_pedestal}
\end{figure}

Finally, the simultaneous response of both detector systems of the EDS was investigated using a linear Compton spectrum produced at FACET-II. Figure~\ref{fig:screen_facet} shows a single-shot image of the beam on the scintillating screen, with the deflection in millimeters determined using equation~\eqref{eq:dipole_deflection}. The \SI{10}{\giga\electronvolt} primary beam is visible at a deflection of about \SI{59}{\milli\meter} as a dark blue region due to saturation of the camera. Electrons that interacted with the laser have lower energies and are therefore deflected further to the left. The corresponding response of the Cherenkov detector is shown in figure~\ref{fig:facet_darkening_pedestal}. The primary beam hits the lowest steel straw, resulting in the large signal peak. In both figures, the vertical dashed lines indicate the range covered by the glass rods. A quantitative comparison of the two detector responses was not possible due to the limited spatial resolution, unresolved Cherenkov-detector backgrounds, and beam fluctuations, and will require further beam-test measurements.

To investigate non-optical background contributions, a shutter between the straws and the SiPMs was used to block the Cherenkov light from reaching the photosensors while leaving the beam--straw interaction unchanged. Figure~\ref{fig:facet_darkening_pedestal} compares the relative signal of all channels with the shutter open and closed, as well as their difference. For steel straws at deflections smaller than the primary beam, the signals with the shutter open and closed are consistent, and their difference is compatible with zero. This indicates that the measured response in this region is dominated by non-optical background. This is expected, as no electrons are present at energies above that of the primary beam. One possible contribution is from beam-induced particles directly interacting with the SiPMs, as already considered for the background observed at ARES, although their individual contributions could not be separated. In contrast, a substantial shutter-dependent response remains for the glass-rod channels. This response does not decrease significantly toward larger deflections as expected from the decreasing electron flux and can therefore not be interpreted solely as Cherenkov light produced by the electron spectrum. One contribution to this background was identified as electronic pickup on the common printed circuit board (PCB) used by the SiPMs of one module~\cite{Klein_aCSPMA_2026}. A signal in one channel of the glass-rod module induced a response in the other channels connected to the same PCB, resulting in an apparent signal even in regions with very low electron flux. Further investigation is required to understand and mitigate this electronic pickup.

\begin{figure}[tbh]
    \centering
    \includegraphics[width=0.70\linewidth]{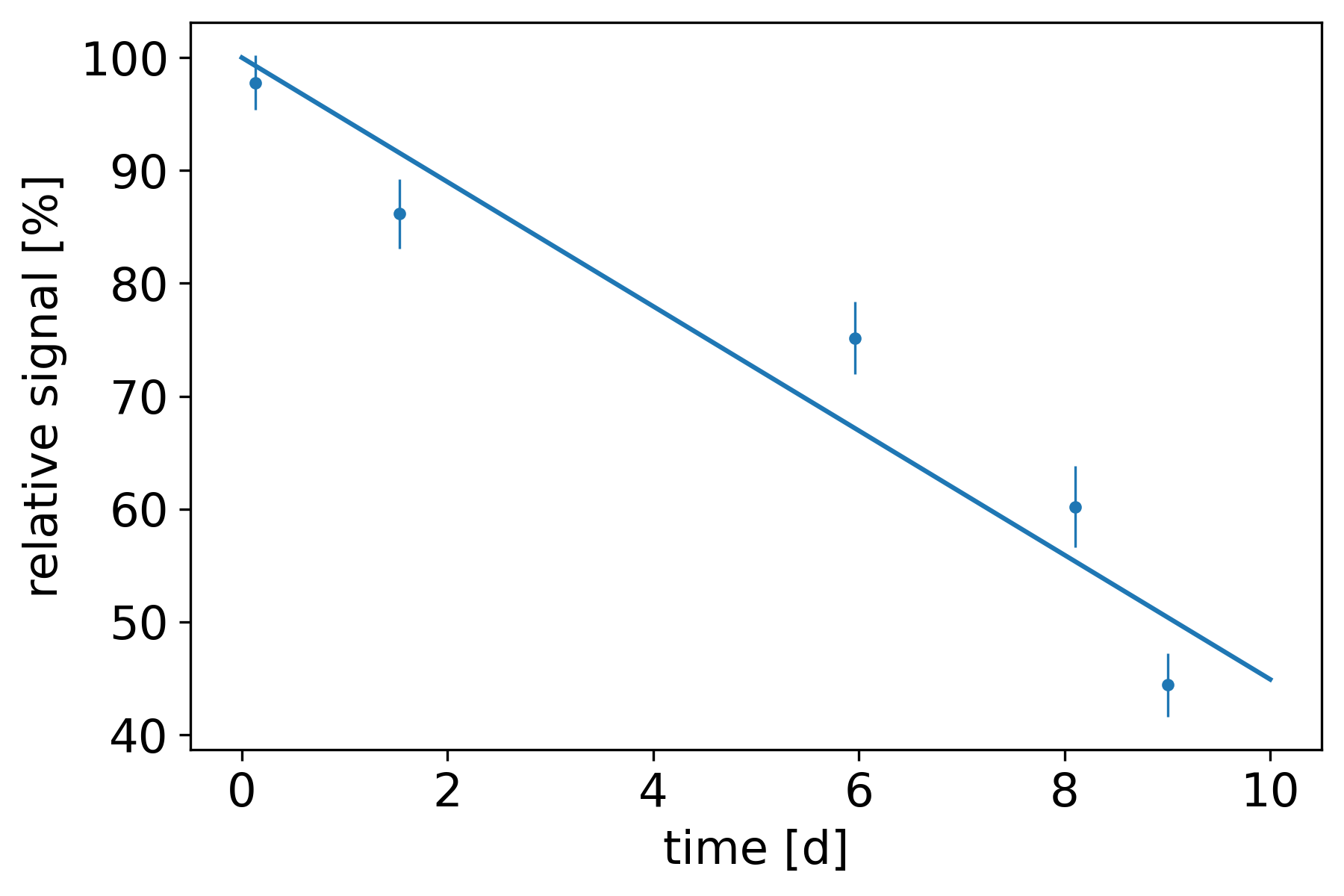}
    \caption{Relative signal intensity from the LED calibration system over the course of nine days. The solid line corresponds to a linear fit to the data. }
    \label{fig:facet_raddamage}
\end{figure}

During the nine days in which the detector was exposed to the running accelerator, a signal intensity reduction of more than 50\% was observed using the LED calibration system (figure~\ref{fig:facet_raddamage}). The EDS was exposed to a high amount of ionizing radiation from the facility's main beam dump, which alters the characteristics of the SiPMs by creating bulk or surface defects in the silicon. These effects can lead to a decrease in gain and photon-detection efficiency, as well as an increase in dark current~\cite{Garutti:2018hfu}. To distinguish between degradation of the SiPMs and radiation-induced losses in the optical fibers used for LED calibration, individual components were replaced after irradiation. At a SiPM bias voltage of \SI{42}{\volt}, replacing an irradiated fiber with a new one increased the measured signal by a factor of \numrange{1.2}{1.5}, whereas replacing a PCB carrying irradiated SiPMs with a new one resulted in an increase by a factor of \numrange{6}{7}. These measurements confirm radiation-induced degradation of both components, with degradation of the SiPMs being the dominant contribution to the observed loss in detector response.

The radiation environment at the European XFEL, where LUXE will be located, is expected to be significantly less severe compared to the conditions near the main beam dump at FACET-II. As a result, degradation of the detector components, particularly the SiPMs, is anticipated to occur at a much slower rate, although the exact impact remains to be investigated in detail.

\section{Conclusion}\label{sec:conclusion}

A high-flux electron detection system is being developed to measure the energy spectrum of electrons produced in laser--electron collisions at the LUXE experiment planned at DESY. The system will combine a scintillating screen with a camera system and a segmented detector based on the Cherenkov effect, using air-filled steel straws and solid glass rods. Prototype tests at the two accelerator facilities ARES at DESY and FACET-II at SLAC demonstrated that the system is advancing toward fulfilling performance requirements for spatial resolution, dynamic range, and radiation tolerance. The screen response was found to be linear and position-independent within the tested range, and the Cherenkov detector showed sensitivity to beam profiles and charge, with calibration techniques successfully applied and several background contributions identified. These results validate the detector concept and provide input for the final design of the system to be used at LUXE. Future work will focus on optimizing the spatial segmentation and improving the sensitivity of the system for measurements at lower beam charges and broader energy spectra. Further beam-test campaigns should also include dedicated measurements to systematically characterize background contributions,  including potential electronic pickup associated with the SiPM PCBs, improve the understanding of the Cherenkov-detector response, and investigate the observed differences in the beam profiles reconstructed with the two detector technologies.

\acknowledgments{The authors acknowledge support from DESY (Hamburg, Germany), a member of the Helmholtz Association HGF. The authors gratefully acknowledge the technical support of Karsten Gadow, Michelle Klotz, Richie Nölling, and Beata Liss at DESY. We also express our sincere gratitude to the E-320 and FACET-II teams at SLAC, including the facility administration and user support staff, for their invaluable remote and on-site support during the preparation and execution of the beam-test campaign. This work was supported by the Swiss National Science Foundation under grants no.\ 214492 and 230596, the European Union’s Horizon Europe Marie Sklodowska-Curie Staff Exchange Program under grant agreement no.\ 101086276 (EAJADE), the Quantum Universe Excellence Cluster, and the University of Hamburg's MIN Graduate Center research travel grant. }

\section*{Conflict of Interest}
The authors have no conflicts of interest to disclose.

\section*{Author Contributions}

Contributions of all authors according to ANSI/NISO~\cite{nisocreditworkinggroup_ansi_2022}. 

\textbf{Conceptualization:} B. Heinemann, L. Helary, R. Jacobs, J. List, M. Wing; 
\textbf{Data curation:} A. Athanassiadis, L. Helary, L. Hendriks, E. Ranken, I. Schulthess;
\textbf{Formal analysis:} A. Athanassiadis, L. Hendriks, I. Schulthess;
\textbf{Funding acquisition:} A. Athanassiadis, L. Hendriks, J. List, E. Ranken, I. Schulthess, M. Wing;
\textbf{Investigation:} A. Athanassiadis, J. Hallford, L. Helary, L. Hendriks, R. Jacobs, E. Ranken, I. Schulthess;
\textbf{Methodology:} A. Athanassiadis, J. Hallford, L. Helary, L. Hendriks, R. Jacobs, J. List, E. Ranken, I. Schulthess, M. Wing;
\textbf{Project administration:} A. Athanassiadis, L. Helary, R. Jacobs, J. List, M. Wing;
\textbf{Resources:} F. Burkart, H. Dinter-Habermeier, L. Helary, R. Jacobs, M. Kellermeier, W. Kuropka, J. List, F. Mayet, S. Schmitt, T. Vinatier, M. Wing;
\textbf{Software:} A. Athanassiadis, O. Borysov, L. Helary, L. Hendriks, F. Mayet, I. Schulthess;
\textbf{Supervision:} F. Burkart, L. Helary, J. List, G. Moortgat-Pick, I. Schulthess, M. Wing;
\textbf{Validation:} A. Athanassiadis, L. Helary, L. Hendriks, R. Jacobs, J. List, E. Ranken, I. Schulthess, M. Wing;
\textbf{Visualization:} A. Athanassiadis, L. Hendriks, I. Schulthess;
\textbf{Writing -- original draft:} A. Athanassiadis, L. Hendriks, I. Schulthess;
\textbf{Writing -- review \& editing:} A. Athanassiadis, O. Borysov, F. Burkart, H. Dinter-Habermeier, J. Hallford, B. Heinemann, L. Helary, L. Hendriks, R. Jacobs, M. Kellermeier, W. Kuropka, J. List, F. Mayet, G. Moortgat-Pick, E. Ranken, S. Schmitt, I. Schulthess, T. Vinatier, M. Wing.


\section*{Data Availability}
The data that support the findings of this study are available from the corresponding author upon reasonable request.

\bibliographystyle{JHEP}
\bibliography{ref.bib}

@article{Athanassiadis:2025git,
    author = "Athanassiadis, Antonios and others",
    title = "{A high-flux electron detector system to measure non-linear Compton scattering at LUXE}",
    eprint = "2505.14720",
    archivePrefix = "arXiv",
    primaryClass = "physics.ins-det",
    doi = "10.1016/j.nima.2025.170777",
    journal = "Nucl. Instrum. Meth. A",
    volume = "1080",
    pages = "170777",
    year = "2025"
}

@article{Abramowicz:2021zja,
    author = "Abramowicz, H. and others",
    title = "{Conceptual design report for the LUXE experiment}",
    eprint = "2102.02032",
    archivePrefix = "arXiv",
    primaryClass = "hep-ex",
    reportNumber = "DESY 21-016, DESY-21-016",
    doi = "10.1140/epjs/s11734-021-00249-z",
    journal = "Eur. Phys. J. ST",
    volume = "230",
    number = "11",
    pages = "2445--2560",
    year = "2021"
}

@article{LUXE:2023crk,
    author = "Abramowicz, H. and others",
    collaboration = "LUXE",
    title = "{Technical Design Report for the LUXE experiment}",
    eprint = "2308.00515",
    archivePrefix = "arXiv",
    primaryClass = "hep-ex",
    doi = "10.1140/epjs/s11734-024-01164-9",
    journal = "Eur. Phys. J. ST",
    volume = "233",
    number = "10",
    pages = "1709--1974",
    year = "2024"
}

@book{Leroy:2011goz,
    author = "Leroy, Claude and Rancoita, Pier-Giorgio",
    title = "{Principles of radiation interaction in matter and detection}",
    doi = "10.1142/5578",
    isbn = "978-981-238-909-1",
    publisher = "World Scientific",
    address = "Singapore",
    year = "2011"
}

@article{Burkart:2022kdx,
    author = "Burkart, Florian and A{\ss}mann, Ralph and Dinter, Hannes and Jaster-Merz, Sonja and Kuropka, Willi and Mayet, Frank and Vinatier, Thomas",
    title = "{The ARES Linac at DESY}",
    doi = "10.18429/JACoW-LINAC2022-THPOJO01",
    journal = "JACoW",
    volume = "LINAC2022",
    pages = "THPOJO01",
    year = "2022"
}

@article{Ritus:1985vta,
    author = "Ritus, V. I.",
    title = "{Quantum effects of the interaction of elementary particles with an intense electromagnetic field}",
    doi = "10.1007/BF01120220",
    journal = "J. Russ. Laser Res.",
    volume = "6",
    pages = "497--617",
    year = "1985"
}

@article{DiPiazza:2011tq,
    author = "Di Piazza, A. and Muller, C. and Hatsagortsyan, K. Z. and Keitel, C. H.",
    title = "{Extremely high-intensity laser interactions with fundamental quantum systems}",
    eprint = "1111.3886",
    archivePrefix = "arXiv",
    primaryClass = "hep-ph",
    doi = "10.1103/RevModPhys.84.1177",
    journal = "Rev. Mod. Phys.",
    volume = "84",
    pages = "1177",
    year = "2012"
}

@article{Gonoskov:2021hwf,
    author = "Gonoskov, A. and Blackburn, T. G. and Marklund, M. and Bulanov, S. S.",
    title = "{Charged particle motion and radiation in strong electromagnetic fields}",
    eprint = "2107.02161",
    archivePrefix = "arXiv",
    primaryClass = "physics.plasm-ph",
    doi = "10.1103/RevModPhys.94.045001",
    journal = "Rev. Mod. Phys.",
    volume = "94",
    number = "4",
    pages = "045001",
    year = "2022"
}

@article{Fedotov:2022ely,
    author = "Fedotov, A. and Ilderton, A. and Karbstein, F. and King, B. and Seipt, D. and Taya, H. and Torgrimsson, G.",
    title = "{Advances in QED with intense background fields}",
    eprint = "2203.00019",
    archivePrefix = "arXiv",
    primaryClass = "hep-ph",
    reportNumber = "RIKEN-iTHEMS-Report-22",
    doi = "10.1016/j.physrep.2023.01.003",
    journal = "Phys. Rept.",
    volume = "1010",
    pages = "1--138",
    year = "2023"
}

@article{Burke:1997ew,
    author = "Burke, D. L. and others",
    title = "{Positron production in multi - photon light by light scattering}",
    reportNumber = "SLAC-PUB-7564, PRINCETON-HEP-97-8, UR-1501, UTKHEP-00010",
    doi = "10.1103/PhysRevLett.79.1626",
    journal = "Phys. Rev. Lett.",
    volume = "79",
    pages = "1626--1629",
    year = "1997"
}

@article{Uggerhj:2005wgn,
    author = "Uggerhj, Ulrik Ingerslev",
    title = "{The interaction of relativistic particles with strong crystalline fields}",
    doi = "10.1103/RevModPhys.77.1131",
    journal = "Rev. Mod. Phys.",
    volume = "77",
    pages = "1131--1171",
    year = "2005"
}

@article{DiPiazza:2019vwb,
    author = "Di Piazza, A. and Wistisen, T. N. and Tamburini, M. and Uggerh{\o}j, U. I.",
    title = "{Testing Strong Field QED Close to the Fully Nonperturbative Regime Using Aligned Crystals}",
    eprint = "1911.04749",
    archivePrefix = "arXiv",
    primaryClass = "hep-ph",
    doi = "10.1103/PhysRevLett.124.044801",
    journal = "Phys. Rev. Lett.",
    volume = "124",
    number = "4",
    pages = "044801",
    year = "2020"
}

@article{Nielsen:2023icv,
    author = "Nielsen, Christian F. and Holtzapple, Robert and Lund, Mads M. and Surrow, Jeppe H. and S{\o}rensen, Allan H. and S{\o}rensen, Marc B. and Uggerh{\o}j, Ulrik I.",
    collaboration = "CERN NA63",
    title = "{Differential measurement of trident production in strong electromagnetic fields}",
    eprint = "2307.03464",
    archivePrefix = "arXiv",
    primaryClass = "hep-ex",
    doi = "10.1103/PhysRevD.108.052013",
    journal = "Phys. Rev. D",
    volume = "108",
    number = "5",
    pages = "052013",
    year = "2023"
}

@misc{Altarelli:2006zza,
    author = "Altarelli, Massimo and others",
    title = "{XFEL: The European X-Ray Free-Electron Laser. Technical design report}",
    reportNumber = "DESY-06-097",
    year = "2006",
    howpublished = "\href{https://doi.org/10.3204/DESY\_06-097}{10.3204/DESY\_06-097}"
}

@article{Kim:2025uhi,
    author = "Kim, Sang Pyo",
    title = "{Strong Field QED, Astrophysics, and~Laboratory Astrophysics}",
    doi = "10.1007/978-981-95-1513-4_2",
    journal = "Springer Proc. Phys.",
    volume = "432",
    pages = "7--12",
    year = "2026"
}

@article{Yokoya:1991qz,
    author = "Yokoya, Kaoru and Chen, Pisin",
    editor = "Dienes, Margaret and Month, Melvin and Turner, Stuart",
    title = "{Beam-beam phenomena in linear colliders}",
    reportNumber = "KEK-PREPRINT-91-2",
    doi = "10.1007/3-540-55250-2_37",
    journal = "Lect. Notes Phys.",
    volume = "400",
    pages = "415--445",
    year = "1992"
}

@article{Yokoya:2000bv,
    author = "Yokoya, K.",
    editor = "Kurokawa, S. I. and Lee, S. Y. and Miles, J. and Perevedentsev, E. A.",
    title = "{Beam-beam interaction in linear collider}",
    doi = "10.1063/1.1420416",
    journal = "AIP Conf. Proc.",
    volume = "592",
    number = "1",
    pages = "185--204",
    year = "2001"
}

@inproceedings{Schulte:1999xb,
    author = "Schulte, D.",
    title = "{High-energy beam-beam effects in CLIC}",
    booktitle={Proceedings of the 1999 Particle Accelerator Conference (Cat. No.99CH36366)}, 
    reportNumber = "CERN-PS-99-017-LP, CERN-PS-99-17-LP, CERN-CLIC-NOTE-391, CLIC-NOTE-391",
    volume = "3",
    year = "1999", 
    doi = "10.1109/PAC.1999.794216"
}

@article{Barklow:2023iav,
    author = "Barklow, Tim and others",
    title = "{Beam delivery and beamstrahlung considerations for ultra-high energy linear colliders}",
    doi = "10.1088/1748-0221/18/09/P09022",
    journal = "JINST",
    volume = "18",
    number = "09",
    pages = "P09022",
    year = "2023"
}

@article{LinearColliderVision:2025hlt,
    author = "Abramowicz, H. and others",
    collaboration = "Linear Collider Vision",
    title = "{A linear collider vision for the future of particle physics}",
    doi = "10.1140/epjs/s11734-026-02153-w",
    journal = "Eur. Phys. J. ST",
    volume = "235",
    number = "6",
    pages = "1641--1796",
    year = "2026"
}

@article{Cole:2017zca,
    author = "Cole, J. M. and others",
    title = "{Experimental evidence of radiation reaction in the collision of a high-intensity laser pulse with a laser-wakefield accelerated electron beam}",
    eprint = "1707.06821",
    archivePrefix = "arXiv",
    primaryClass = "physics.plasm-ph",
    doi = "10.1103/PhysRevX.8.011020",
    journal = "Phys. Rev. X",
    volume = "8",
    number = "1",
    pages = "011020",
    year = "2018"
}

@article{Yakimenko:2019sya,
    author = "Yakimenko, V. and others",
    title = "{FACET-II facility for advanced accelerator experimental tests}",
    doi = "10.1103/PhysRevAccelBeams.22.101301",
    journal = "Phys. Rev. Accel. Beams",
    volume = "22",
    number = "10",
    pages = "101301",
    year = "2019"
}

@article{Garutti:2018hfu,
    author = "Garutti, E. and Musienko, Yu.",
    title = "{Radiation damage of SiPMs}",
    eprint = "1809.06361",
    archivePrefix = "arXiv",
    primaryClass = "physics.ins-det",
    doi = "10.1016/j.nima.2018.10.191",
    journal = "Nucl. Instrum. Meth. A",
    volume = "926",
    pages = "69--84",
    year = "2019"
}

@article{Kropf:2025loq,
    author = "Kropf, Annabel and Schulthess, Ivo",
    title = "{Primer of Strong-Field Quantum Electrodynamics for Experimentalists}",
    doi = "10.3390/physics8010026",
    journal = "MDPI Physics",
    volume = "8",
    number = "1",
    pages = "26",
    year = "2026"
}

@article{Nakamura:2011zzc,
    author = "Nakamura, K. and Gonsalves, A. J. and Lin, C. and Smith, A. and Rodgers, D. and Donahue, R. and Byrne, W. and Leemans, W. P.",
    title = "{Electron beam charge diagnostics for laser plasma accelerators}",
    doi = "10.1103/PhysRevSTAB.14.062801",
    journal = "Phys. Rev. ST Accel. Beams",
    volume = "14",
    pages = "062801",
    year = "2011"
}

@article{glinec_absolute_2006,
  title = {Absolute Calibration for a Broad Range Single Shot Electron Spectrometer},
  author = {Glinec, Y. and Faure, J. and {Guemnie-Tafo}, A. and Malka, V. and Monard, H. and Larbre, J. P. and De Waele, V. and Marignier, J. L. and Mostafavi, M.},
  year = 2006,
  month = oct,
  journal = {Review of Scientific Instruments},
  volume = {77},
  number = {10},
  pages = {103301},
  issn = {0034-6748, 1089-7623},
  doi = {10.1063/1.2360988},
  urldate = {2026-09-24},
  langid = {english},
}

@misc{DRZ,
    title = {{DRZ Screens}},
    author = {{MCI Optonix}},
    url= {https://web.archive.org/web/20251008005808/https://mcio.com/products/drz-screens.aspx},
    urldate = {2025-10-15},
    howpublished = {\url{https://web.archive.org/web/20251008005808/https://mcio.com/products/drz-screens.aspx}},
    year = {2025}
}

@misc{hamamatsu_photonics_kk_hamamatsu_2023,
	title = {Hamamatsu {MPPC} S14160 Datasheet},
	url = {https://www.hamamatsu.com/content/dam/hamamatsu-photonics/sites/documents/99_SALES_LIBRARY/ssd/s14160-1310ps_etc_kapd1070e.pdf},
	author = {{HAMAMATSU PHOTONICS K.K.}},
	urldate = {2024-03-23},
    howpublished = {\url{https://www.hamamatsu.com/content/dam/hamamatsu-photonics/sites/documents/99_SALES_LIBRARY/ssd/s14160-1310ps_etc_kapd1070e.pdf}},
	year = {2023},
}

@misc{nisocreditworkinggroup_ansi_2022,
  title = {{{ANSI}}/{{NISO Z39}}.104-2022, {{CRediT}}, {{Contributor Roles Taxonomy}}},
  author = {{NISO CRediT Working Group}},
  year = {2022},
  month = feb,
  publisher = {NISO},
  urldate = {2025-06-11},
  langid = {english},
  howpublished = {\href{https://doi.org/10.3789/ansi.niso.z39.104-2022}{10.3789/ansi.niso.z39.104-2022}}
}

@article{Danson:2019qlu,
    author = "Danson, Colin N. and others",
    title = "{Petawatt and exawatt class lasers worldwide}",
    doi = "10.1017/hpl.2019.36",
    journal = "High Power Laser Sci. Eng.",
    volume = "7",
    pages = "e54",
    year = "2019"
}

@misc{reisE320ProgressFY242024,
  title = {E-320 {{Progress}} in {{FY24}} and {{Plans}} for {{FY25}}},
  author = {Reis, David and Meuren, Sebastian},
  year = 2024,
  address = {SLAC, US},
  howpublished = {\url{https://indico.slac.stanford.edu/event/9280/contributions/10855/attachments/4822/12981/2024_NOV_E320_FACET.pdf}},
  langid = {english},
}

@article{lindhard_motion_1964,
  title = {Motion of Swift Charged Particles, as Influenced by Strings of Atoms in Crystals},
  author = {Lindhard, J.},
  year = 1964,
  month = sep,
  journal = {Physics Letters},
  volume = {12},
  number = {2},
  pages = {126--128},
  issn = {00319163},
  doi = {10.1016/0031-9163(64)91133-3},
  urldate = {2025-10-20},
  copyright = {https://www.elsevier.com/tdm/userlicense/1.0/},
  langid = {english}
}

@article{opencv_library,
    author = {Bradski, G.},
    journal = {Dr. Dobb's Journal of Software Tools},
    title = {{The OpenCV Library}},
    year = {2000}
}

@article{Blackburn:2023mlo,
    author = "Blackburn, T. G. and King, B. and Tang, S.",
    title = "{Simulations of laser-driven strong-field QED with Ptarmigan: Resolving wavelength-scale interference and {\ensuremath{\gamma}}-ray polarization}",
    doi = "10.1063/5.0159963",
    journal = "Phys. Plasmas",
    volume = "30",
    number = "9",
    pages = "093903",
    year = "2023"
}

@misc{ptarmigan_github_2024,
    author = "Blackburn, T. G. and Fleck, K.",
    title = "tgblackburn/ptarmigan (version v1.4.2) [Computer software]",
    howpublished = "\href{https://doi.org/10.5281/zenodo.14169754}{10.5281/zenodo.14169754}",
    year = "2024"
}

@phdthesis{Keeble:2019eqo,
    author = "Keeble, Fearghus Robert",
    title = "{Measurement of the electron energy distribution at AWAKE}",
    school = "University Coll. London",
    year = "2019"
}

@article{Bauche:2019vjt,
    author = "Bauche, J. and others",
    title = "{A magnetic spectrometer to measure electron bunches accelerated at AWAKE}",
    eprint = "1902.05752",
    archivePrefix = "arXiv",
    primaryClass = "physics.ins-det",
    doi = "10.1016/j.nima.2019.05.067",
    journal = "Nucl. Instrum. Meth. A",
    volume = "940",
    pages = "103--108",
    year = "2019"
}

@misc{2KCam,
    title = {{Basler ace GigE camera acA1920-40gm}},
    author = {{Basler AG}},
    url= {https://docs.baslerweb.com/aca1920-40gm},
    urldate = {2025-11-01},
    howpublished = {\url{https://docs.baslerweb.com/aca1920-40gm}},
    year = {2025}
}

@misc{4KCam,
    title = {{Basler ace GigE camera aca4096-11gm}},
    author = {{Basler AG}},
    url= {https://docs.baslerweb.com/aca4096-11gm},
    howpublished = {\url{https://docs.baslerweb.com/aca4096-11gm}},
    urldate = {2025-11-01},
    year = {2025}
}

@misc{dejong_caenv1730daq_2023,
  title = {{{CAEN-v1730-DAQ}}},
  author = {{de Jong}, Sam},
  year = {2023},
  howpublished={\url{https://github.com/samdejong86/CAEN-v1730-DAQ}},
  month = jun
}

@misc{caenspa_ds3153_2019,
  title = {{{DS3153}} - 730 {{Digitizer Family}} 16/8 {{Channel}} 14-Bit 500 {{MS}}/s {{Data Sheet}}},
  author = {{CAEN SpA}},
  year = {2019},
  month = nov,
  urldate = {2024-09-25},
  howpublished = {\url{https://caen.it/products/v1730-v1730s/}},
  langid = {english},
}

@misc{caenspa_ds3159_2019,
  title = {{{DS3159}} - 742 {{Digitizer Family}} 32+2/16+1 {{Channel}} 12-Bit 5 {{GS}}/s {{Switched Capacitor Data Sheet}}},
  author = {{CAEN SpA}},
  year = 2019,
  month = dec,
  howpublished = {\url{https://caen.it/products/v1742/}},
  urldate = {2025-11-06},
  langid = {english},
}

@inproceedings{Hensler:1996ppq,
    author = "Hensler, O. and Rehlich, K.",
    title = "{DOOCS: a Distributed Object Oriented Control System}",
    booktitle = "{15th Conference on Charged Particle Accelerators}",
    pages = "308--315",
    year = "1996"
}

@misc{epicscontrols_experimental_2024,
  title = {The {{Experimental Physics}} and {{Industrial Control System EPICS}}},
  author = {{EPICS Controls}},
  year = 2024,
  howpublished = {\url{https://github.com/epics-base/epics-base}}
}

@article{Gessner:2021wgi,
    author = "Gessner, Spencer",
    title = "{The FACET-II Data Acquisition System}",
    doi = "10.18429/JACoW-IBIC2021-WEPP33",
    journal = "JACoW",
    volume = "IBIC2021",
    pages = "WEPP33",
    year = "2021"
}

@book{stefanov_cmos_2022,
  title = {{{CMOS Image Sensors}}},
  shorttitle = {{{CMOS Image Sensors}}},
  author = {Stefanov, Konstantin D},
  year = 2022,
  month = nov,
  publisher = {IOP Publishing},
  doi = {10.1088/978-0-7503-3235-4},
  urldate = {2025-11-10},
  isbn = {978-0-7503-3235-4},
}

@article{HERNANDEZADAME20188,
title = {Effect of Tb3+ concentration in the visible emission of terbium-doped gadolinium oxysulfide microspheres},
journal = {Solid State Sciences},
volume = {84},
pages = {8-14},
year = {2018},
issn = {1293-2558},
doi = {https://doi.org/10.1016/j.solidstatesciences.2018.07.021},
url = {https://www.sciencedirect.com/science/article/pii/S1293255818303960},
author = {Luis Hernandez-Adame and Gabriela Palestino and Octavio Meza and Pablo Luis Hernandez-Adame and Hector Rene Vega-Carrillo and Iyad Sarhid},
}

@incollection{AHMED2015435,
title = {7 - Position-sensitive detection and imaging},
editor = {Syed Naeem Ahmed},
booktitle = {Physics and Engineering of Radiation Detection (Second Edition)},
publisher = {Elsevier},
edition = {Second Edition},
pages = {435-475},
year = {2015},
isbn = {978-0-12-801363-2},
doi = {https://doi.org/10.1016/B978-0-12-801363-2.00007-3},
url = {https://www.sciencedirect.com/science/article/pii/B9780128013632000073},
author = {Syed Naeem Ahmed},
}

@article{Schwinkendorf:2019zrz,
    author = "Schwinkendorf, J. P. and others",
    title = "{Charge calibration of DRZ scintillation phosphor screens}",
    doi = "10.1088/1748-0221/14/09/P09025",
    journal = "JINST",
    volume = "14",
    number = "09",
    pages = "P09025",
    year = "2019"
}

@misc{Liu:2026nln,
    author = "Liu, Shuang and others",
    title = "{Absolute charge calibration of DRZ phosphor screens for relativistic electron bunches}",
    eprint = "2607.17059",
    archivePrefix = "arXiv",
    primaryClass = "physics.acc-ph",
    howpublished = "\href{https://arxiv.org/abs/2607.17059}{2607.17059}",
    year = "2026"
}

@article{Lensch:2023wgx,
    author = "Lensch, Timmy and Lipka, Dirk and Neumann, Reinhard and Werner, Matthias",
    title = "{Comparison of Different Bunch Charge Monitors Used at the ARES Accelerator at DESY}",
    doi = "10.18429/JACoW-IBIC2023-TU3I04",
    journal = "JACoW",
    volume = "IBIC2023",
    pages = "TU3I04",
    year = "2023"
}

@mastersthesis{Klein_aCSPMA_2026,
    author = "Klein, D.",
    title = "{aCSPMA}: a {Cherenkov} {Silicon} {Photomultiplier} {Array}",
    school = "University of Hamburg",
    year = "2026"
}

\end{document}